\documentclass{amsart}
\usepackage{graphicx}
\newtheorem{thm}{Theorem}[section]
\newtheorem{cor}[thm]{Corollary}
\newtheorem{lem}[thm]{Lemma}

\theoremstyle{definition}
\newtheorem{defn}[thm]{Definition}
\theoremstyle{remark}
\newtheorem{rem}[thm]{Remark}
\numberwithin{equation}{section}

\newlength{\uppermar}
\newlength{\lowermar}
\newlength{\leftmar}
\newlength{\rightmar}

\usepackage[centertags]{amsmath}
\usepackage{amsfonts}
\usepackage{amssymb}
\usepackage{amsthm}
\usepackage[english]{babel}
\usepackage{calc}

\begin{document}

\title[Characteristic classes of gauge systems]{Characteristic classes of gauge systems}
\author{S.L.~Lyakhovich and A.A.~Sharapov}
\address{Department of Quantum Field Theory, Tomsk State University, Tomsk 634050, Russia}
\email{sll@phys.tsu.ru, sharapov@phys.tsu.ru}

\thanks{The authors are indebted to I.A. Batalin, V.A. Dolgushev and M.A. Grigoriev for useful
discussions on various issues related to this paper. This work
benefited from the following research grants: INTAS grant 00-262,
RFBR grant 03-02-17657, Russian Ministry of Education grant E
02-3.1-250, and the grant for Support of Russian Scientific
Schools 1743.2003.2. AAS appreciates the financial support from
the Dynasty Foundation and the International Center for
Fundamental Physics in Moscow.}




\begin{abstract}
We define and study invariants which can be uniformly constructed
for any gauge system. By a gauge system we understand an
(anti-)Poisson supermanifold endowed  with an odd Hamiltonian
self-commuting vector field called a homological vector field.
This definition encompasses all the cases usually included into
the notion of a gauge theory in physics as well as some other
similar (but different) structures like Lie or Courant algebroids.
For Lagrangian gauge theories or Hamiltonian first class
constrained systems, the homological vector field is identified
with the classical BRST transformation operator. We define
characteristic classes of a gauge system as universal cohomology
classes of the homological vector field, which are uniformly
constructed in terms of this vector field itself. Not striving to
exhaustively classify all the characteristic classes in this work,
we compute those invariants which are built up in terms of the
first derivatives of the homological vector field. We also
consider the cohomological operations in the space of all the
characteristic classes. In particular,  we show that the
(anti-)Poisson bracket becomes trivial when applied to the space
of all the characteristic classes, instead the latter space can be
endowed with another Lie bracket operation. Making use of this Lie
bracket one can generate new characteristic classes involving
higher derivatives of the homological vector field. The simplest
characteristic classes are illustrated by the examples relating
them to anomalies in the traditional BV or BFV-BRST theory and to
characteristic classes of (singular) foliations.

\end{abstract}
\maketitle
\section{Introduction}

Most of the fundamental physical models, especially those from the
field theory, are the gauge systems in the traditional sense, i.e.
the models whose action functionals are invariant under the local
(gauge) transformations of the configuration space.
 At the level of local geometry,
 the conventional gauge theories were well studied in seventies and
eighties in the framework of the BRST-BV-BFV approach
\cite{BV1,BV2,BV3,BFV2,BFV3,HT} which had been initially designed
as a universal method for quantizing these theories. With time
this method has evolved into the theory having much broader range
of physical and mathematical applications including computation of
anomalies, constructing consistent interactions of gauge fields,
quantizing wide classes of Poisson manifolds related with Lie
algebroids, etc.

The BRST method works with a gauge system by constructing its
special super-extension in such a way as to convert the gauge
invariance into a special global symmetry generated by an odd
vector field $\mathcal{Q}$ (a generator of the BRST
transformation). Being one-dimensional and odd, the corresponding
Lie algebra is necessarily Abelian,
\begin{equation}
[\mathcal{Q},\mathcal{Q}]=0\,. \label{Q20}
\end{equation}
The physical observables of a conventional gauge theory are then
identified with certain cohomology classes of the BRST
differential $\mathcal{Q}$, signifying the fundamental role this
operator plays in physics. In mathematics, a growing trend is
observed to study the Lie algebroids and the Courant algebroids
through their embedding into an extended target superspace
equipped with the nilpotent odd vector field, called a homological
vector field \cite{Vain,Roy,Vor}. Though the homological
description of the algebroids is somewhat different from the BRST
theory of traditional gauge systems, both the types of gauge
theories can be put into the uniform homological scheme. In this
paper we hold a wider understanding of the gauge system (the
accurate definition is given in the next section) encompassing
both traditional models of field theory and a range of other
problems. All the cases covered have in common that the original
manifold of the theory can be embedded into a supermanifold
endowed with a nilpotent odd vector field $\mathcal{Q}$ which
``absorbs'' the algebraic structure behind the original system. In
doing so, the distinctions between different types of original
gauge systems are encoded in the ghost number grading prescription
for the extended supermanifold.

The global properties of the gauge systems remain yet much less
studied even though the BRST methods offer promise as a processing
tool for that. One of the obstacles to studying the global
geometry of gauge systems by the BRST tools is that the BRST
embedding procedure, as it was originally proposed, is not
explicitly covariant with respect to natural automorphisms of the
gauge algebra:  the structure functions entering the BRST
generator are neither tensors nor connections, that makes the
covariance of the BRST embedding not so evident. This difficulty
is not crucial, however, for the BRST theory and, as we show
below, it can be easily overcome by introducing an appropriate
connection on the bundle of the ghost variables.

Once the covariant formulation is established, various questions
can be addressed concerning the global geometry of the gauge
systems, e.g.: Given the gauge system, to which simplest form it
can be brought to by a suitable automorphism? Given two gauge
systems, defined on the same supermanifold, are they equivalent to
each other? In the local setting, the exhaustive answers to these
questions are provided by the Abelization theorem \cite{BFR, BVc}
which can be regarded as a generalization of the Darboux theorem
to the case of the (anti-)symplectic supermanifolds endowed with a
regular Poisson nilpotent superfunction (BRST charge or master
action, depending on the parity of the Poisson bracket). In
general, we cannot expect to find a globally defined Abelian basis
for the gauge algebra generators as the existence of this basis
imposes very strong restrictions on the global geometry. In the
finite dimensional case, for instance, the gauge orbits should
cover a torus to make the system globally abelianizable. In the
field theory, dealing with infinite-dimensional manifolds, the
very definition of a global equivalence represents a problem, not
to mention  the in-depth study of the topology of the
infinite-dimensional gauge orbits. This suggests to look for other
topological invariants which could be both easily computable and
sufficiently informative to perceive locally invisible differences
between gauge systems.

In this paper, we begin studying the characteristic classes which
reflect the global properties of the gauge systems. The
characteristic classes are defined by the cocycles of the
homological vector field which are explicitly built up in terms of
this operator itself. In the particular case of Lie algebroids
some of our characteristic classes seem to coincide with the
characteristic classes introduced by Fernandes \cite{F}, and, in
the more special situation of an abstract Lie algebra, they are
reduced to the primitive generators of the Lie algebra cohomology.
For other types of gauge systems falling into this uniform scheme,
no general expressions were known before for the characteristic
classes, as far as we know.

The paper is organized as follows. In Sec. \ref{DE} we formulate
the general notion of the gauge system and consider some important
examples, demonstrating how different types of physical and
geometrical models can be brought into the uniform framework of
the gauge system theory. Some of the considered examples contain
new results, e.g. we propose a new Hamiltonian gauge embedding for
the Lie algebroids, for the BV and BFV formalisms we give slightly
modified, explicitly covariant formulations introducing a
connection in the corresponding ghost bundles. In Sec. \ref{cgs}
we discuss the cohomology of gauge systems with tensor
coefficients. It is argued that the tensor cohomology can be
reduced, in a sense, to the scalar one by an appropriate extension
of the original gauge system. The opposite operation, i.e. the
reduction on an invariant submanifold, gives rise to additional
cohomological invariants which can be related with a gauge system.
The notions of the universal cocycles and characteristic classes
are introduced in Sec. \ref{HC}. In this section we also
explicitly construct a simplest series of characteristic classes.
Sec. \ref{COP} is devoted to natural cohomological operations
which can be introduced on the space of all characteristic
classes. We show that the characteristic classes are Poisson
commuting, so the Poisson bracket induces a trivial cohomological
operation among the topological invariants; instead, the linear
space of all characteristic classes carries a special Lie bracket
operation, which  can be used, for example, to generate new
characteristic classes from the already known ones. Sec. \ref{apl}
is devoted to the interpretation of simplest characteristic
classes as one-loop quantum anomalies in the BV and BRST-BFV
formalism as well as to application of these classes to the theory
of (singular) foliations. In concluding section we briefly
summarize the results and discuss some prospects of further
studies. Appendix collects basic formulas and conventions on the
differential geometry of supermanifolds used throughout the text.

\section{Gauge systems: definition and examples.}\label{DE}

In this paper we work in the category of  the supermanifolds
endowed with an even or odd Poisson structure\footnote{In the
physical literature the odd Poisson bracket is usually called
anti-bracket.}. Recall that the Poisson structure on a
supermanifold $M$ is a bilinear map $\{\cdot,\cdot\}:
C^\infty(M)\otimes C^\infty(M)\to C^\infty(M)$, called bracket,
satisfying the following axioms:
\begin{equation}\label{PB}
    \begin{array}{ll}
    \epsilon(\{f,g\})=\epsilon(f) +\epsilon(g)+\epsilon&  ({\rm {mod}}\,\; 2) \,,\\[3mm]
      \{f,g\}=-\{g,f\}(-1)^{(\epsilon (f)+\epsilon)(\epsilon(g)+\epsilon)} &
      (symmetry)\,, \\[3mm]
      \{f,gh\}=\{f,g\}h+\{f,h\}g (-1)^{\epsilon(g)\epsilon(h)}& (Leibnitz\;\;rule)\,,
      \\[3mm]
       \{f,\{g,h\}\}(-1)^{(\epsilon (f)+\epsilon)(\epsilon(h)+\epsilon)}+cycle(f,g,h)=0&
       (Jacobi\;\; identity)\, ,
    \end{array}
\end{equation}
where $\epsilon= \epsilon(\{\,\cdot,\cdot\,\})$ is the Grassman
parity of the bracket ($\epsilon=0$ for the even Poisson
structure, and $\epsilon=1$ for the odd one). The pair
$(M,\{\cdot,\cdot\})$ is called a Poisson manifold.

Hereinafter we omit the prefix ``super'' whenever possible, e.g.
the terms like manifold, vector bundle, matrix, function, etc.
will actually mean the corresponding notions of super-mathematics.

Now, let us specify what we understand by the gauge system.

\begin{defn} A \textit{gauge  system} is  a triple
$(M,\{\cdot,\cdot\},Q)$ constituted by a smooth manifold $M$,
endowed with a Poisson bracket $\{\,\cdot,\cdot\,\}$  and a
function $Q\in C^{\infty}(M)$, such that
\begin{equation}\label{gs}
     \epsilon (Q)=\epsilon (\{\cdot,\cdot\}) + 1 \,,\qquad\{Q,Q\}=0\,.
\end{equation}
\end{defn}

\noindent As the parity of $Q$ is opposite to the parity of the
bracket, the last relation does not hold identically, it is a
nontrivial condition imposed on the function $Q$. From (\ref{gs})
also follows that the Hamiltonian vector field
$\mathcal{Q}=\{Q,\cdot\}$ is always odd and integrable
\begin{equation}\label{hvf}
\epsilon(\mathcal{Q}) =1\,,\;\;\;\;\;\;
[\mathcal{Q},\mathcal{Q}]=2\mathcal{Q}^2=0\,.
\end{equation}

In this paper we adopt the following terminology. Any odd and
nilpotent vector field is called a \textit{homological vector
field}. In the case of a Hamiltonian homological vector field
$\mathcal{Q}=\{Q,\cdot\}$, the Hamiltonian $Q$ is referred to as a
\textit{ homological potential}. The condition (\ref{gs}) on the
homological potential is called a \textit{master equation} (odd or
even, depending on the parity of the Poisson bracket involved).

The morphisms can be obviously defined in the category of gauge
systems: Given two gauge systems $\mathcal{G} =
(M,\{\,\cdot,\cdot\,\},Q)$ and
$\mathcal{G'}=(M',\{\,\cdot,\cdot\,\}', Q')$, a smooth map
$\varphi: M\to M'$ defines a morphism
$\mathcal{G}\mapsto\mathcal{G'}$ if
\begin{equation}\label{morph}
\varphi^\ast
(\{f,g\}')=\{\varphi^\ast(f),\varphi^\ast(g)\}\,,\qquad
\varphi^\ast (Q')=Q\,,\quad \forall \, f,g\in C^{\infty}(M')\,.
\end{equation}
In other words, the morphisms are just Poisson maps relating
homological potentials. When $\varphi$ is a diffeomorphism,  the
gauge systems $\mathcal{G}$ and $\mathcal{G'}$ are said to be
equivalent.

One can also pick out several suitable subcategories in the
category of all gauge systems. Given a homological vector field
$\mathcal{Q}=\mathcal{Q}^i\partial_i$ and an affine connection
$\nabla$, introduce the $(1,1)$-tensor field
\begin{equation}\label{Hesse}
    \Lambda =(\nabla_i \mathcal{Q}^j)\,.
\end{equation}
If $p\in M$ is a stationary point of $\mathcal{Q}$, i.e.
$\mathcal{Q}^i(p)=0$ then, treating $\Lambda(p)=\Lambda_p$ as the
matrix of an odd linear operator $\Lambda_p: T^\ast_pM\to
T^\ast_pM$, one can see that
\begin{equation}\label{L2}
\Lambda_p^2=0\,.
\end{equation}
As $\Lambda_p$ is a nilpotent operator, one can define the
corresponding cohomology group
\begin{equation}\label{Hp}
H_p=\ker \Lambda_p/\mathrm{im} \Lambda_p.
\end{equation}
 The set of all stationary points of the
homological vector field $\mathcal{Q}$ is called a
\textit{stationary shell} and denoted $\Sigma =\{p\in
M|\mathcal{Q}^i(p)=0\}$.

\begin{defn}
 A gauge system $(M,\{\,\cdot,\cdot\,\},Q)$ is said to be
\textit{regular} if the cohomology group $H_p$ has the same
    dimension at any stationary point:
$$
    \dim H_p=const\,,\qquad\forall p\in \Sigma\,.
$$
\end{defn}

\rem The regularity condition ensures $\Sigma\subset M$ to be a
smooth submanifold. The usual notion of a corank of the Jacobi
matrix associated to the system of equations $\mathcal{Q}^i(p)=0$,
is not directly applicable to the odd matrix $\Lambda_p$ and the
dimension of the cohomology group $H_p$ substitutes this notion,
in a sense.

The local structure of a regular gauge system was established by
different methods in Refs. \cite{BV3, BFR, BVc, S1, S2, AKSZ}. (In
fact, these works addressed either BV theory or the BRST-BFV one,
although their results hold true for other types of the gauge
systems as well). Given a regular gauge system $(M,\{\cdot ,\cdot
\},Q)$, then a coordinate system $(x^i, \xi^i, \eta^\alpha)$
exists in a neighbourhood of every point $p\in \Sigma$ , such that
\begin{equation}\label{}
\epsilon(\xi^i)=\epsilon(x^i)+1 \,,\qquad \alpha=1,..., \dim
H_p\,,
\end{equation}
and the homological vector field has the form
\begin{equation}\label{}
    \mathcal{Q}=\xi^i\frac{\partial}{\partial x^i}\,.
\end{equation}
In these adapted coordinates the stationary shell $\Sigma$ is
determined by the equations $\xi^i =0$. The geometry of the
stationary shell has been further studied in \cite{GST1, GST2}
using both the BRST-BFV and the BV formalism.

Another important subcategory of gauge systems is related to the
notion of $\mathbb{Z}$-grading, traditionally called in physics
the ghost-number grading. In mathematics, $\mathbb{Z}$-graded
supermanifolds are characterized by introducing an additional
integer grading in the structure sheaf \cite{Roy}, \cite{Vor}.
Simply stated, a $\mathbb{Z}$-graded manifold $M$ is a
supermanifold possessing an atlas of affine charts in which each
local coordinate is assigned an integer number called weight and
the admissible coordinate transformations respect the total
weight. It is appropriate mention that the supplementary
$\mathbb{Z}$-grading is in no way related to the underlying
$\mathbb{Z}_2$-grading (the Grassman parity) of a supermanifold.
In what follows we prefer to use the physical terminology adopted
in the BRST theory and refer to $\mathbb{Z}$-grading as the
\textit{ghost grading}; accordingly, the weights of local
coordinates $x^i$, denoted by $\mathrm{gh}(x^i)$, will be called
the \textit{ghost numbers}.

Notice that on the graded manifold $M$ all geometric objects
(tensor fields, connections, etc.) carry ghost numbers, which can
be conveniently described by means of the Euler vector field
\begin{equation}\label{gnumber}
\hat{G}=\sum_i \mathrm{gh}(x^i)x^i\frac{\partial}{\partial x^i} \,
, \qquad \mathrm{gh}(x^i) \in \mathbb{Z} \, .
\end{equation}
The ghost numbers are just the eigenvalues of the operator of the
Lie derivative along $\hat{G}$: the tensor field $S$ on $M$ is
said to have a ghost number $n\in\mathbb{Z}$ if
$\mathcal{L}_{\hat{G}} S=nS$. For coordinate functions on a local
chart this gives the definition of $\mathrm{gh} (x^i)$.

The stationary points of the Euler vector field  $\hat{G}$ form a
smooth submanifold $N\subset M$, which we will call the
\textit{body} of $M$. According to this definition the local
coordinates on $N$ carry zero ghost number. In physics, it is the
body $N$ which is usually considered as ``an original'' space
which is then ``BRST-embedded'' into the super-extension $M$.

In the graded category, the homological vector field $\mathcal{Q}$
is usually required to carry the ghost number $+1$, i.e.
\begin{equation}\label{ghq}
    [\hat{G},\mathcal{Q}]=\mathcal{Q}\,,
\end{equation}
and this results in the following correlation between ghost
numbers of the homological potential and the bracket:
\begin{equation}\label{}
    \mathrm{gh} (Q)=1-\mathrm{gh}(\{\,\cdot,\cdot\,\})\,.
\end{equation}
It is the Lie superalgebra (\ref{hvf}), (\ref{ghq}), spanned by
two generators $\mathcal{Q}$ and $G$, which is known as the
BRST-algebra.

The cohomology group (\ref{Hp}) associated to the stationary shell
$\Sigma\subset M$ inherits the ghost grading from $M$: If the
spectrum of ghost numbers ranges form $-n$ to $m$, then $$H_p =
\bigoplus_{k=-n}^m H_p^k\,.$$ Besides the notion of regularity, an
important notion of the properness of a gauge system can be
formulated in this case.

\begin{defn}  A regular gauge  system is said to be a \textit{proper}
if
\begin{equation}\label{prop}
H_p^k=0\,,\qquad \forall k < 0\,.
\end{equation}
\end{defn}

For instance, the properness condition (\ref{prop}) is
automatically satisfied if there are no coordinates with negative
ghost numbers on $M$. In the conventional BRST theory it is the
properness condition which provides the unique existence of the
solution to the master equation with a given boundary condition.

\subsection{Examples.} Here we exemplify the general definitions by reviewing some
known constructions which have been intensively studied both in
physics and mathematics, and which all fit into the concept of the
gauge system explained above. The basic characteristics of a gauge
system  are the spectrum of ghost numbers, the parity and ghost
number of the homological potential $Q$ (that automatically
defines the ghost number and parity of the bracket). Except of the
case of Hamiltonian first-class systems, a $\mathbb{Z}$-graded
manifold $M$ underlying a gauge system involves the cotangent
bundle $T^\ast N$ to the body of $M$, the fibers of which carry a
non-zero ghost number\footnote{The more precise description of a
gauge system includes also the specification  of the ghost number
and parity distribution among all the coordinates on $M$.}. To get
a general impression how different types of gauge systems are
embedded in the uniform homological framework, it is convenient to
arrange the corresponding data in the table:

\vspace{2mm}\noindent
\begin{tabular}{|c|c|c|c|c|c|}

\hline

{}&{} & {} & {} & {} &{}\\[-2pt]


\;\;N\;\;& Type of the gauge system & Spectrum of  ghost numbers&$\; \mathrm{gh} ( Q)\; $& $\epsilon (Q)$&$\mathrm{gh}(T_p^\ast N)$\\[6pt]

  \hline

{}&{} & {} & {} & {} &{}\\[-2pt]

 1.& Irreducible Lagrangian gauge system & $-2, \; -1, \; 0 \; ,1$ &  0 & 0 &-1\\[6pt]

  \hline

{}&{} & {} & {} & {} &{}\\[-2pt]

 2.&$n$-times reducible Lagrangian gauge system  & $-2 - n, \dots, n+1$ & 0 & 0& -1 \\[6pt]

  \hline

{}&{} & {} & {} & {}&{} \\[-2pt]

3.& Hamiltonian first class constrained system &$ -1, \;0, \; 1 $& 1 & 1& -- \\[6pt]

  \hline

{}&{} & {} & {} & {}&{} \\[-2pt]

 4.&$n$-times reducible first class  sytem& $-n-1, \dots, n+1 $ &1 & 1 & --\\[6pt]

  \hline

{}&{} & {} & {} & {} &{}\\[-2pt]
5.&Lie algebroid & $0, \; 1,\; 2,\; 3$ & 4 & 1 & 3\\[6pt]

  \hline

{}&{} & {} & {} & {} &{}\\[-2pt]

6.& $n$-times reducible Lie algebroid & $0, \dots, 2n+1 $ & $2n+2$ & 1& $2n+1$ \\[6pt]

  \hline

{}&{} & {} & {} & {}&{} \\[-2pt]

7.&Courant algebroid & $0,\; 1, \; 2 $& 3 & 1& 2 \\[6pt]

  \hline

{}&{} & {} & {} & {}&{} \\[-2pt]

8.&Weak Poisson bracket & $-1, \; 0, \;1, \;   2$ & 2 & 0 & 1\\[6pt]

 \hline

\end{tabular}

\vspace{0.5cm}

\noindent Examples 1, 3 are well known in physics and we present
them below in a conventional way. A slightly novel point here is
that we use explicitly covariant (even or odd) Poisson brackets
involving the connection to describe the gauge models both in the
Hamiltonian and Lagrangian formalisms\footnote{Similar brackets
for the ghosts have been found earlier by I.A.Batalin,
M.A.Grigoriev and one of us (SLL) for another type of the gauge
system as a part of a different work which still remains
unfinished. We also truly appreciate Maxim Grigoriev's comments on
the relevance of the connection in the ghost bundle for this
construction.}. Examples 2, 4 are the extensions of the previous
two ones to the case of reducible gauge algebra generators
\cite{BV2}, \cite{BFV3}. Although the Lie algebroid may appear,
for instance, as a particular solution to the BV-master equation
of the Lagrangian gauge theory, the latter solution would be
neither proper nor invariant w.r.t. automorphisms of the BRST
(\ref{gs}), (\ref{ghq}). The genuine BRST-like imbedding for the
Lie algebroid requires another spectrum of ghost numbers,
indicated in fifth line of the table, and we elaborate on this
example below. The case of $n$-times reducible Lie algebroid
(Example 6), extends the previous one in the same sense as 2 and 4
extend 1 and 3 respectively. This case will be considered
elsewhere, here we just mention the grading of the appropriate
manifold to embed the reducible algebroid into the master
equation. Example 7 is due to Roytenberg \cite{Roy}, who has first
translated the Courant algebroid into the language of the master
equation. The last example concerns the embedding of a Poisson
manifold into an almost-Poisson manifold so that the imbedding map
to relate the corresponding bivectors \cite{WeakDB}.

\subsubsection{BV anti-field formalism.}\label{lagr}
The Lagrangian description of a field-theoretical model with an
irreducible gauge symmetry is implemented in terms of a
$\mathbb{Z}$-graded manifold $M$, equipped with an odd symplectic
structure. Topologically, $M$ is given by a direct sum of vector
bundles\footnote{Hereinafter, the numbers in square brackets point
on the ghost number of the fiber coordinates.} $\Pi T^\ast[-1]
N\oplus \Pi \mathcal{E}[1]\oplus \mathcal{E}^\ast[-2]$, where $N$
is a configuration space (more accurately, the space of
trajectories) of the original gauge invariant model and
$\mathcal{E}\to N$ is a vector bundle whose sections are the
parameters of the gauge transformations.  Local coordinates on the
original configuration space (fields, in physics) are denoted
$\phi^A$. Linear coordinates on the fibres of $\Pi T^\ast[-1] N$,
$\Pi \mathcal{E}[1]$ and $\mathcal{E}^\ast[-2]$ are, respectively,
$\phi^\ast_A$ (anti-fields), $C^\alpha$ (ghosts) and
$C^\ast_\beta$ (anti-ghosts). By definition,
\begin{equation}\label{gr}
\begin{array}{llll}
\epsilon(\phi^\ast_A)=\epsilon_A
+1\,,\;\;\;\;\;\;&\epsilon(C^\alpha)=\epsilon_\alpha+1\,,\;\;\;\;\;\;&
\epsilon(C^\ast_\alpha)=\epsilon_\alpha\,,
\;\;\;\;\;&({\rm mod}\, 2)\,,\\[3mm]
    \mathrm{gh}(\phi^A)=0\,,\;\;\;\;\;&
    \mathrm{gh}(\phi^\ast_A)=-1\,,\;\;\;\;\;&  \mathrm{gh}(C^\alpha
    )=1\,,\;\;\;\;\;&
    \mathrm{gh}(C^\ast_\beta)=-2\,,
    \end{array}
\end{equation}
where $\epsilon_A=\epsilon (\phi^A)$,
$\epsilon_\alpha=\epsilon(\varepsilon^\alpha) $, and
$\varepsilon^\alpha$ are the parameters of the gauge
transformations. Below, for the sake of simplicity the original
configuration space $N$ is assumed to be an ordinary (even)
 manifold.

Upon choosing a linear connection $\nabla$ on $\mathcal{E}\to N$
one can endow $M$ with the structure of an odd Poisson manifold.
The nonvanishing Poisson brackets among the local coordinates are
\begin{equation}\label{obr}
\begin{array}{c}
  \{\phi^\ast_B, \phi^A\}=\delta_B^A\,,\;\;\;\;\;\;\;\{C^\ast_\beta, C^\alpha\}=\delta_\beta^\alpha\,,
  \\[3mm]
   \{\phi^\ast_A,C^\alpha\}=\Gamma_{A\,\beta}^\alpha(\phi)
   C^\beta\,,\;\;\;\;\;\;\;\;\;
   \{\phi^\ast_A,C^\ast_\beta\}=-\Gamma_{A\,\beta}^\alpha(\phi)C_\alpha\,,\\[3mm]
   \{\phi^\ast_A,\phi^\ast_B\}=R_{AB}{}^\alpha_{\beta}(\phi) C^\beta
   C^\ast_\alpha\,,
\end{array}
\end{equation}
where $\Gamma_{A\beta}^\alpha(\phi)$ and
$R_{AB}{}^\alpha_\beta(\phi)$ are components of the connection
$\nabla$ and its curvature tensor, respectively.  The Poisson
bracket (\ref{obr}) increases the ghost number by 1, $$\textrm{gh}
(\{f,g \}) = \textrm{gh}(f) + \textrm{gh}(g) +1, \,\quad \forall
f,g \in \mathcal{C}^\infty(M)\,,$$ and comes from the following
(exact) odd symplectic structure\footnote{We omit the sign of the
wedge product treating the differentials $dx^i$ as basis sections
of $\Pi T^\ast M$ (see Appendix A).}
\begin{equation}\label{oss}
\omega=d(\phi^\ast_Ad\phi^A+C^\ast_\alpha \nabla C^\alpha)
=d\phi^\ast_A d\phi^A+\nabla C^\ast_\alpha \nabla
C^\alpha+\frac12d\phi^A d\phi^B R_{AB}{}_\alpha^\beta  C^\alpha
C^\ast_\beta\,,
\end{equation}
$$
\nabla C^\alpha=dC^\alpha +d\phi^A\Gamma_{A\beta}^\alpha
C^\beta\,,\qquad \nabla
C^\ast_\beta=dC^\ast_\beta-d\phi^A\Gamma_{A\beta}^\alpha
C^\ast_\alpha\,.
$$

Considering $M$ as an abstract $\mathbb{Z}$-graded manifold it is
natural to extend the group of coordinate transformations (which
are originally linear in the fiber coordinates $\phi^\ast, C$ and
$C^\ast$) to an arbitrary smooth coordinate changes respecting the
Grasmann parity and the ghost-number grading. In this case,
different choices for the connection $\nabla$ will lead to
equivalent symplectic structures on $M$. Namely, any two
symplectic structures $\omega$ and $\omega'$ of the form
(\ref{oss}) are related to each other by the coordinate transform
$\phi^\ast_A\to \phi^\ast_A+\Delta
\Gamma_{A\alpha}^\beta(\phi)C^\alpha C^\ast_\beta$, where
$\Delta\Gamma=\nabla-\nabla'$ is the difference between
corresponding connections. Moreover, one can prove an odd
counterpart of Rothstein's theorem \cite{R}, stating that any odd
symplectic structure of the ghost number 1 can be brought into the
form (\ref{obr}) by a suitable diffeomorphism of $M$.

A solution to the master equation (\ref{gs}), denoted usually by
$S$, is called a \textit{master action}. As the odd bracket
carries the ghost number 1, $\mathrm{gh}(S)=0$. The corresponding
homological vector field $\mathcal{Q}=\{S,\,\cdot\,\}$ is known as
the \textit{generator of the BRST transformation}. The properness
condition can be written as
\begin{equation}\label{bc}
 \mathrm{ rank} \,(d\,{}^2S)|_{dS=0}={\dim \mathcal{E}}=\frac 12\dim M\,,
\end{equation}
where $d^2S=(\partial_i\partial_jS)$ is the Hesse matrix and the
equation $dS(\phi)=0$ is supposed to have a solution.

With the account of ghost numbers and parities of the local
coordinates we can write
\begin{equation}\label{ma}
    S(x)={\textsf{S}}(\phi)+C^\alpha R_\alpha^A(\phi)\phi^\ast_A+\frac12
    C^\alpha C^\beta T_{\beta\alpha}^\gamma(\phi) C^\ast_\gamma
    +\frac14 C^\beta C^\alpha
    E^{AB}_{\alpha\beta} (\phi)\phi^\ast_A\phi^\ast_B + O(C^3)\,.
\end{equation}
Substituting this expansion into the odd master equation
\begin{equation}
\label{BVme} \{S,S\}=0\,,
\end{equation}
one can find all the higher order ghost and anti-field terms
provided (\ref{BVme}) is satisfied  up to the second order in
$C$'s \cite{BV3,HT}. In this lower order in ghosts the equation
(\ref{BVme}) is reduced to the relations
\begin{equation}\label{ga}
R_\alpha^A\partial_A \textsf{S}
=0\,,\;\;\;\;\;\;\;\;[R_\alpha,R_\beta]=U_{\alpha\beta}^\gamma
R_\gamma +\partial_A {\textsf{S}} \,
E^{AB}_{\alpha\beta}\partial_B\,,
\end{equation}
where
$$U_{\alpha\beta}^\gamma=T_{\alpha\beta}^\gamma-R_\alpha^A
\Gamma_{A\beta}^\gamma+R_\beta^A\Gamma_{A\alpha}^\gamma\,.$$ The
first relation means that the function $\textsf{S}(\phi)$,
identified with the action function of a gauge model, is invariant
under the action of the (local) vector fields
$R_\alpha=R_\alpha^A\partial_A$, while the second relation implies
these vector fields to form an integrable distribution on the
surface $\Sigma'\subset N:\; d{\textsf{S}}=0$.  Notice that
$\Sigma'$ is the projection on $N$ of the stationary  shell
$\Sigma: dS=0$.  In the physical literature the relations
(\ref{ga}) are called a \textit{gauge algebra}. Accounting the
rank condition (\ref{bc}), one can see that (i) the local vector
fields $R_\alpha$ are lineally independent on $\Sigma'$, and (ii)
the corank of the Hesse matrix $d^2{\textsf{S}}$ on $\Sigma'$ is
equal to the ${\rm rank}\, \mathcal{E}$, so that $\Sigma'\subset
N$ is a smooth surface indeed.

Applying the standard  technique of the cohomological perturbation
theory \cite{HT} it is possible to prove the inverse: given the
classical action function $\textsf{S}\in C^\infty (N)$ with the
Hesse matrix $d^2\textsf{S} $ having a constant rank on $\Sigma'$,
there exists a unique (up to symplectomorphisms of $M$) proper
solution to the Batalin-Vilkovisky master equation (\ref{BVme})
starting with $\textsf{S}$.

We see that any gauge algebra (\ref{ga}) defines a Lie algebroid
$\mathcal{E}|_{\Sigma'}$ over the stationary surface $\Sigma'$
with the anchor $R|_{\Sigma'}: \mathcal{E}|_{\Sigma'}\rightarrow
T\Sigma'$. The action of the algebroid foliates $\Sigma'$ onto the
gauge orbits. Mention that the gauge algebra does not necessarily
define a Lie algebroid structure beyond $\Sigma'$ (and hence it
defines no foliation structure in a tubular neighbourhood of
$\Sigma'\subset N$) because of  possible contributions of the
\textit{trivial gauge generators} (second term in the  r.h.s. of
(\ref{ga})). In physical literature, this more general structure,
with an  admixture of the trivial gauge generators (vanishing on
the equations of motion $d{\textsf{S}}=0$) in the r.h.s., is
called an \textit{open algebra}. The BV theory encodes all the
structure relations of this more complicate algebra by means of a
single master equation (\ref{BVme}) supplemented with the
properness condition (\ref{bc}).

\subsubsection{BRST-BFV Hamiltonian formalism.} Consider BFV
description of the Hamiltonian systems subject to irreducible
first class constraints \cite{BFV2}, \cite{BFV3}, \cite{HT}. The
description involves a graded manifold $M$  associated to the
total space of the direct sum $\Pi\mathcal{E}[1]\oplus
\Pi\mathcal{E}^\ast[-1]$, where $\mathcal{E}\to N$ is a vector
bundle over a symplectic manifold $(N,\omega^0)$. Let $x^i$ denote
local coordinates on the base $N$, and let $C^\alpha$ and
$P_\alpha$ denote linear coordinates on fibres of $\Pi\mathcal{E}$
and $\Pi\mathcal{E}^\ast$, respectively. In physical literature
$C$'s and $P$'s are called the ghost coordinates and momenta (or
just ghosts). According to the definition of $M$,
 \begin{equation}\label{bfvc}
\mathrm{gh}(x^i)=0\,,\quad \textrm{gh} (C^\alpha) = 1 \, , \quad
\textrm{gh}(P_\alpha) = - 1 \, .
\end{equation}
Again, to simplify  the exposition, we assume $N$ to be an even
manifold, so that $\epsilon(x^i)=0$,
$\epsilon(C^\alpha)=\epsilon(P_\alpha)=1$. Using a linear
connection $\nabla$ on $\mathcal{E}\rightarrow N$ one can extend
the even symplectic structure $\omega^0$ from $N$ to $M$ as
follows:
\begin{equation}\label{bfvo}
\omega= dx^i dx^j\omega^0_{ij} + d (P_\alpha \nabla C^\alpha) =
dx^i dx^j\omega^0_{ij} + \nabla P_\alpha \nabla C^\alpha+ \frac12
dx^i dx^jR_{ij}{}_{\alpha}^\beta   C^\alpha P_\beta \,,
\end{equation}
$$\nabla C^\alpha = d C^\alpha + dx^i\Gamma_i{}^{\alpha}_{\beta}C^\beta\, , \qquad
 \nabla P_\alpha = d P_\alpha -
dx^i\Gamma_i{}_{\alpha}^{\beta} P_\beta \,,$$ where
$\Gamma_i{}_{\alpha}^{\beta}(x)$, and $R_{ij}{}_{\alpha}^\beta
(x)$ are the components of the connection $\nabla$ and its
curvature tensor. The corresponding  Poisson bracket on $M$ reads
\begin{equation}\label{bfvbr}
\begin{array}{ll}
\{x^i,x^j\}=\Pi^{ij}\,,&\{C^\alpha,P_\beta\}=\delta_\beta^\alpha-\Pi^{ij}
\Gamma_{i\mu}^\alpha\Gamma_{j\beta}^\nu C^\mu P_\nu\,,\\[3mm]
    \{x^i,C^\alpha\}=\Pi^{ij}\Gamma_{j\beta}^\alpha C^\beta\,,&
    \{x^i,P_\beta\}=-\Pi^{ij}\Gamma _{j\beta}^\alpha P_\alpha\,, \\[3mm]
     \{C^\alpha , C^\beta\}= \Pi^{ij}\Gamma_{i\mu}^\alpha \Gamma_{j\nu}^\beta C^\mu  C^\nu\,,\qquad&
\{P_\alpha , P_\beta\}= \Pi^{ij}\Gamma_{i\alpha}^\mu
\Gamma_{j\beta}^\nu P_\mu  P_\nu\,,
    \end{array}
\end{equation}
where
\begin{equation}\label{}
    \Pi^{ij}=\pi^{ik}\sum_{\textrm{m}=0}^{\infty} (R^\textrm{m})_k^j \, , \qquad
    R=(R_j^i)=\left(\frac 12 \pi^{ik}R_{kj}{}_\alpha^\beta
    C^\alpha P_\beta\right)\,,
\end{equation}
and $\pi$ is the Poisson bivector dual to the symplectic 2-form
$\omega^0$ on $N$. An adaptation of Rothstein's theorem \cite{R}
to the setting of $\mathbb{Z}$-graded manifolds shows that any
Poisson structure on $M$ is equivalent to that given by
(\ref{bfvbr}).

Mention that, unlike the previous example, the bracket does not
change the ghost number, i.e. $\textrm{gh} (\{f,g \}) =
\textrm{gh}(f) + \textrm{gh}(g)$, and the ghost number operator
$\hat{G}$ can be realized in the adjoint manner
\begin{equation}\label{bfvgh}
\hat{G}f =\{G, f\} = \mathrm{gh}(f)f \,, \qquad G = C^\alpha
P_\alpha \,,
\end{equation}
for any homogeneous function $f\in C^{\infty}(M)$.  Since the
bracket is even, an odd homological potential $\Omega$ is to be
introduced. In the BRST-BFV theory $\Omega$ carries the ghost
number 1, and is called the BRST charge. So, one has
\begin{equation}\label{bfvme}
 \{\Omega,\Omega\}=0\,, \qquad \{ G, \Omega  \} = \Omega \, .
\end{equation}
The last relations are known as the BRST algebra. Since $M$ is a
symplectic manifold the properness  condition for the generator of
the BRST  transformations $\mathcal{Q}=\{\Omega,\,\cdot\,\}$ can
be written as
\begin{equation}\label{regcon1}
    \mathrm{rank}(d\,{}^2\Omega)|_{d\Omega=0}=2\,\mathrm{rank}\,\mathcal{E}\,,
\end{equation}
where $d\,{}^2\Omega =(\partial_A\partial_B \Omega)$ is an even
matrix if one changes the parity of the first index.   Accounting
the ghost number distribution, the general expression for the BRST
charge reads
\begin{equation}\label{bfvq}
\Omega= C^\alpha T_\alpha (x) + \frac12 C^\alpha C^\beta
T_{\beta\alpha}^\gamma (x)P_\gamma + O(P^2)\,.
\end{equation}
The coefficients $T_\alpha(x)$ at the first order in ghosts are
identified with the Hamiltonian constraints on the original phase
space $N$. The properness condition (\ref{regcon1}) allows to
conclude that equations $T_\alpha (x)=0$ single out a smooth
submanifold (the constraint surface) $\Sigma'\subset N$, being the
projection on $N$ of the stationary shell $\Sigma
=\Pi\mathcal{E}^\ast[-1]|_{\Sigma'}$.

In the lowest order in $C$s', the master equation (\ref{bfvme})
gives the involution relations for the first class constraints
\begin{equation}\label{invol}
\{ T_\alpha, T_\beta \}_0 = U_{\alpha\beta}^\gamma T_\gamma\,,
\end{equation}
where the Poisson bracket is defined by the bivector $\pi$, and
\begin{equation}\label{}
    U_{\alpha\beta}^\gamma=T_{\alpha\beta}^\gamma -
    \pi^{ij}\partial_iT_\alpha\Gamma_{j\beta}^\gamma +
    \pi^{ij}\partial_iT_\beta\Gamma_{j\alpha}^\gamma\,.
\end{equation}
In other words, $\Sigma'\subset N$ is a coisotropic submanifold
\cite{Wein}.

Again, using a sort of homological perturbation theory \cite{HT},
\cite{St} one can see that (i) all the higher orders in
(\ref{bfvq}) can be iteratively restored provided $\Sigma\subset
N$ is a smooth coistropic submanifold, and (ii) any two solutions
for $\Omega$, associated with the same $\Sigma$, are related to
each other by a simplectomorphism of $M$ \cite{HT}, \cite{St}.

\subsubsection{Lie algebroids}\label{liealg} Let $\mathcal{E}\to N$ be a Lie
algebroid over a smooth manifold $N$ with the anchor $A:
\mathcal{E}\to TN$. Define the graded supermanifold $M=
T^\ast[3]N\oplus \Pi\mathcal{E}[1]\oplus \Pi\mathcal{E}^\ast[2]$.
By the definition,
\begin{equation}\label{}
    \begin{array}{cccc}
      \epsilon(x^i)=0\,,\quad & \epsilon(p_i)=0\,,\quad &
       \epsilon(C^\alpha)=1\,,\quad &
       \epsilon(P_\alpha)=1\,,\\[3mm]
      \mathrm{gh}(x^i)=0\,,\quad & \mathrm{gh}(p_i)=3\,,\quad &
      \mathrm{gh}(C^\alpha)=1\,,\quad & \mathrm{gh}
      (P_\alpha)=2\,.
    \end{array}
\end{equation}
Here $x^i$ are local coordinates on $N$, while  $p_i$, $C^\alpha$
and $P_\alpha$ are linear coordinates on the fibers of
$T^\ast[3]N$, $\Pi \mathcal{E}[1]$ and $\Pi\mathcal{E}^\ast [2]$,
respectively.

Choosing a linear connection $\nabla$ on $\mathcal{E}\to N$ the
manifold $M$ can be endowed with an even symplectic structure of
the ghost number 3. The corresponding Poisson brackets of local
coordinates read
\begin{equation}\label{liebr}
    \begin{array}{c}
      \{p_i,x^j\}=\delta_i^j\,,\quad \{P_\alpha,
      C^\beta\}=\delta_\alpha^\beta\,,\\[3mm]
       \{p_i, C^\beta\}=\Gamma_{i\alpha}^\beta C^\alpha\,,\quad \{p_i, P_\beta\}
       =-\Gamma_{i\beta}^\alpha P_\alpha\,,\\[3mm]
\{p_i, p_j\}=R_{ij}{}_\beta^\alpha C^\beta P_\alpha\,,
\end{array}
\end{equation}
and the other brackets vanish. Here $R_{ij}{}_\alpha^\beta$ is the
curvature tensor of the connection $\nabla=\partial +\Gamma$.
Since the even bracket is introduced with the ghost number -3, the
homological potential $Q$ should be odd and carrying the ghost
number 4. Taking into account the ghost number and parity of the
homological potential on this manifold, the most general
admissible expression for $Q$ reads
\begin{equation}\label{lieq}
Q=C^\alpha A_\alpha^i(x) p_i +\frac12C^\alpha C^\beta
T_{\beta\alpha}^\gamma(x) P_\gamma \,.
\end{equation}
The usual Lie algebroid relations  for the anchor
$A_\alpha^i\in\Gamma(\mathcal{E}^\ast\otimes TN)$ and the torsion
$T_{\alpha\beta}^\gamma\in \Gamma (\mathcal{E}\otimes
\mathcal{E}\wedge \mathcal{E})$ immediately follow from the master
equation $\{Q,Q\}=0$ with the bracket (\ref{liebr}). Notice that
our BRST-like imbedding for the Lie algebroid differs from that
considered in \cite{Roy1}, where the ghost numbers are assigned in
a different way and more than one ghost grading is imposed.

\subsubsection{Courant algebroids}
Consider the manifold $M$ given by the direct sum of vector
bundles $T^\ast[2]N\oplus \Pi\mathcal{E}[1]$ over an even manifold
$N$. If $x^i$ are local coordinates on $N$, and $p_i$ and $\xi^a$
are local coordinates on fibers of $T^\ast[2]N$ and
$\Pi\mathcal{E}$, then we have
\begin{equation}\label{}
\begin{array}{ccc}
\epsilon(x^i)=0\,,\quad&\epsilon(p_i)=0\,,\quad&\epsilon(\xi^a)=1\,,\\[3mm]
    \mathrm{gh} (x^i)=0\,,\quad&
    \mathrm{gh}(p_i)=2\,,\quad&\mathrm{gh} (\xi^a)=1\,.
    \end{array}
\end{equation}
Any symplectic structure on $M$, having  the ghost number 2 is
equivalent to the following one:
\begin{equation}\label{}
  \omega = d(p_idx^i+\xi^ag_{ab}\nabla \xi^b)=dp_idx^i+
   g_{ab}(x)\nabla\xi^a\nabla\xi^b
   +\frac12dx^idx^j R_{ijab}(x)\xi^b \xi^a\,,
\end{equation}
where $\nabla$ is a linear connection on $\mathcal{E}$ respecting
a pseudo-Euclidean metric $g_{ab}$ and having  the curvature
$R_{ijab}=g_{ca}R_{ij}{}^c_b$. The corresponding Poisson brackets
of local coordinates read
\begin{equation}\label{pb2}
\begin{array}{lll}
  \{x^i,x^j\}=0 \,,& \{\xi^a,x^i\}=0\,,  &\{p_i,
  \xi^a\}=\Gamma_{ib}^a(x)\xi^b\,,\\[3mm]
   \{p_i,x^j\}=\delta_i^j\,, \;\;\;\;&
   \{\xi^a,\xi^b\}=\frac12g^{ab}(x)\,,\;\;\;\;
   & \{p_i,p_j\}= R_{ijab}(x) \xi^a\xi^b\,.
\end{array}
\end{equation}

 The most general expression for the homological potential,
 compatible with the ghost number grading and parity, is
given by
\begin{equation}\label{}
    Q=\xi^a A_a^i(q)p_i -\frac 16
    \phi_{abc}(q) \xi^a \xi^b \xi^c\,,\quad\quad \mathrm{gh} (Q)=3\,,
\end{equation}
where $A\in \Gamma (\mathcal{E}^\ast\otimes TN)$ and $\phi \in
\Gamma (\wedge^3\mathcal{E}^\ast)$. Given a connection $\nabla$ on
$N$, the master equation $\{Q,Q\}=0$ imposes certain relations on
the structure functions $g, A$ and $\phi$, which prove to be
equivalent to the structure relations of the Courant algebroid
\cite{C}, \cite{Roy}, ( see also \cite{LWX}) upon identification
of $A$ as the respective anchor.

\section{Cohomology of a gauge system}\label{cgs}

Any homological vector field $\mathcal{Q}$ gives rise to the
nilpotent operator $\delta : \mathcal{T}(M) \to \mathcal{T}(M)$ on
the space of smooth tensor fields:
\begin{equation}\label{delta}
\delta S = \mathcal{L}_\mathcal{Q}S\,,\quad\quad \forall \, S\in
\mathcal{T}(M)\,,
\end{equation}
$\mathcal{L}_Q$ being the Lie derivative along $\mathcal{Q}$. Let
$Z_\delta(\mathcal{T}(M))={\rm ker}\, \delta \,$  to denote the
group of $\delta$-cocycles and let $B_\delta (\mathcal{T}(M))={\rm
im}\, \delta \, $ be the group of $\delta$-coboundaries.  The
$\delta$-cohomology group is the quotient
\begin{equation}
H_\delta(\mathcal{T}(M))=Z_\delta(\mathcal{T}(M))/B_\delta
(\mathcal{T}(M)\,.
\end{equation}
Clearly, the subgroup $Z_\delta (\mathrm{Vect}(M))$ has the
distinguished element $\mathcal{Q}$. When $\mathcal{Q}$ is a
Hamiltonian vector field, one has an additional pair of cocycles:
the homological potential $Q\in Z_\delta(C^\infty(M))$ and the
Poisson bivector $\pi \in Z_{\delta}(\wedge^2 \mathrm{Vect}(M))$.

Since $\delta$ is essentially the generator of an infinitesimal
diffeomorphism,  applying various tensor operations \footnote
{like tensor multiplication, permutation and contraction of
indices, Schouten brackets of polyvector fields, concomitant of
symmetric contravariant tensors, exterior derivative of
differential forms, etc.} to $\delta$-cocycles one can induce the
corresponding operations in the $\delta$-cohomology. Again, in the
case of the gauge systems one has distinguished cohomological
operations e.g. taking the Poisson bracket of two scalar cocycles
or multiplying a tensor cocycle by the Poisson bivector.

In what follows we will mostly deal with the scalar cocycles and
their cohomology groups  $H_\delta (C^\infty(M))=H_\delta(M)$. The
reason is that any tensor cocycle can be always realized as a
scalar cocycle on an appropriately extended manifold $M'$.
Moreover, it is always possible to convert the corresponding
homological vector field $\mathcal{Q}'$ into a Hamiltonian one by
further extending $M'$. Indeed, one can always identify
$\mathcal{T}^{(n,m)}(M)$ with the space of the polylinear
functions on the total space $M'$ of $T^\ast M^{\otimes n}\otimes
TM^{\otimes m}$. The homological vector field $\mathcal{Q}$ is
then lifted to $M'$ by usual formulas of tensor calculus (i.e. as
the Lie derivative). At the next step one can lift $\mathcal{Q}'$
to the Hamiltonian vector field $Q''$ on $T^\ast M'$ (or $\Pi
T^\ast M'$) w.r.t. the canonical symplectic structure:
\begin{equation}\label{}
\begin{array}{rl}
    Q''(x,p)=\mathcal{Q}'^i(x)p_i \,,&
    Q''(x,x^\ast)=Q'^i(x)x^\ast_i\,,\\[3mm]
       \{x^i,p_j\}=\delta_j^i\,,&\{x^i,x^\ast_j\}'=\delta_j^i\,,
\end{array}
\end{equation}
where $x^i$ are the local coordinates on $M'$, while $p_i$ and
$x^\ast_i$ are the linear coordinates on the fibers of $T^\ast M'$
and $\Pi T^\ast M'$, respectively. Thus, studying the general
structure of the homological algebra associated with nilpotent
$\mathcal{Q}$, we can always restrict consideration to the scalar
cocycles.

One may regard the above procedure as an extension of the
homological vector filed $\mathcal{Q}$ from $M$ to $M'$ (of
course, there can be many other procedures achieving the same
goal). Let us describe an inverse procedure, namely, the reduction
of a homological vector field to a submanifold. Given a set of
smooth functions $\{f_\alpha\}$ on the manifold $M$ equipped with
a honological vector field  $\mathcal{Q}$, consider the ideal
$I\subset C^{\infty}(M)$ generated by the extended set of
functions $(f_\alpha,\delta f_\beta) = (\phi_A)$. The equations
$\phi_A=0$ extract a smooth submanifold $N\subset M$ provided the
standard regularity condition is satisfied:
\begin{equation}\label{}
    \mathrm{rank}\,\left(\frac{\partial \phi_A}{\partial
    x^i}\right)_N= const\,.
\end{equation}
By construction, $N$ is the $\mathcal{Q}$-invariant submanifold
endowed with the homological vector field
$\mathcal{Q}'=\mathcal{Q}|_N$. Thus, to any $\delta$-closed ideal
of functions $I$ satisfying the regularity condition, we can
associate a cohomology group $H_{\delta'} (\mathcal{T}(N))$
together with the natural homomorphism
$H_\delta(\mathcal{T}^{(0,\bullet)}(M))\to H_\delta
(\mathcal{T}^{(0,\bullet)}(N))$ induced by the restriction of
covariant tensor fields on the submanifold $N$. In algebraic
terms, the commutative algebra of functions $C^\infty(N)$ can be
equivalently described as the quotient $C^\infty(M)/I$.
Respectively, the group $H_\delta(N)$ is identified with
cohomology classes of relative cocycles: $c\in C^{\infty}(M)$
defines an element $[c]\in H_\delta (N)$ iff $ \delta c\in I$.

One can generalize this construction by replacing $f_\alpha$ with
a set of tensor fields $S_\alpha$ and their
$\delta$-differentials. Again, under the regularity condition, the
equations $S_\alpha=\delta S_\alpha =0$ define a smooth
submanifold equipped with a homological vector field. An important
example of an invariant submanifold is the stationary shell
$\Sigma\subset M$ of a homological vector field $\mathcal{Q}$.

Existence of the invariant submanifolds provides in that way an
additional source for cohomological invariants one can relate with
the gauge system.

In the category of $\mathbb{Z}$-graded manifolds the group
$H_\delta(M)$ is naturally graded with respect to the ghost
number. In field theory one usually deals with the scalar
cohomology of the ghost numbers 0, 1 and 2. The group
$H^0_\delta(M)$ is identified with the \textit{space of physical
observables} of a gauge system, while $H^1_\delta(M)$ is
considered as a source for possible \textit{anomalies} in the
Lagrangian quantization. In the Hamiltonian description the same
quantum anomalies are nested in $H_\delta^2(M)$. We refer to
\cite{BBH}, \cite{BBH1} for general results on  the local BRST
cohomology as well as its application to the Yang-Mills type
theories. In Sec. \ref{apl} we will put the problem of anomalies
in the general context of characteristic classes to be developed
in the next section.

\section{Characteristic classes}\label{HC}
In this section we consider scalar cocycles which can be
universally defined for any gauge system
$(M,\{\,\cdot,\,\,\cdot\}, Q)$. For this reason we call them
\textit{universal cocycles}. The formal definition is as follows.

\begin{defn} Given a Poisson manifold $(M,\{\,\cdot,\,\cdot\})$, a
local map $U :C^{\infty}(M)\to C^{\infty}(M)$ is said to be a
\textit{universal cocycle} if for any $\Theta \in C^{\infty}(M)$
with $\epsilon (\Theta)=\epsilon (\{\,\cdot,\,\cdot\})+1$ there
exists a linear differential operator $\hat{U }:C^{\infty}(M)\to
C^{\infty}(M)$ (depending perhaps on $\Theta$) such that
\label{d1}
\begin{equation}\label{U}
    \{\Theta, U(\Theta)\}=\hat{U }\{\Theta,\Theta\}\,.
\end{equation}
\end{defn}

 The mentioned property of locality means that
\begin{equation}\label{}
    {\rm supp}\; U(\Theta)\subset {\rm supp}\; \Theta\,,\quad\quad \forall \, \Theta\in
    C^\infty(M)\,.
\end{equation}
When $M$ is compact, this amounts to saying that $U(\Theta)$ is
specified by some symmetric polydifferential operator (i.e.
symmetric w.r.t. permutations of its arguments, which all are then
substituted by one and the same function $\Theta \in
C^{\infty}(M)$).

The equation (\ref{U}) merely says that once a homological
potential $Q$ is substituted into $U$ instead of an arbitrary
$\Theta$, the function $U(Q)$ becomes a $\delta$-cocycle. Note
that Definition \ref{d1} involves the Poisson structure underlying
a gauge system but not the homological potential. Taking another
homological potential on the same Poisson manifold, one can find
its $\delta$-cocycles making use of the same map $U$. That is why
we name $U(Q)$ a \textit{universal cocycle}.

By \textit{characteristic classes} of a gauge system we mean the
$\delta$-cohomology classes of its universal cocycles. These form
a subgroup $UH_\delta(M)\subset H_\delta(M)$ in the full group of
$\delta$-cohomology. The natural grading  in the space of
polydifferential operators induces the grading in $UH_\delta(M)$:
\begin{equation}\label{}
    UH^n_\delta(M)\ni [U]\;\;\;\Leftrightarrow\;\;\;
    U(tQ)=t^nU(Q)\,,\qquad t\in \mathbb{R}\,.
\end{equation}
For example, $U(Q)=Q^n$ is the universal cocycle of degree $n$.
Less trivial examples of universal cocycles can be obtained as
follows. Let $\omega\in \Omega^n(M)$ be a closed $n$-form on $M$,
put
\begin{equation}\label{ex}
    U(Q)=i_{\mathcal{Q}}^n\omega\,,
\end{equation}
where $i_{\mathcal{Q}}$ denotes the contraction of a differential
form with the homological vector field $\mathcal{Q}$. It is easy
to check that (\ref{ex}) is a $\delta$-cocycle.  Indeed,
accounting that $i_\mathcal{Q}$ is an even operator we can write
\begin{equation}\label{}\begin{array}{c}
    \delta U(Q)=\mathcal{L}_\mathcal{Q} i_\mathcal{Q}^n\omega
    =-i_\mathcal{Q}d\,i_\mathcal{Q}^n\omega
    = - (i_\mathcal{Q} \mathcal{L}_\mathcal{Q}i^{n-1}_\mathcal{Q}+i^2_\mathcal{Q}d\,i^{n-1}_\mathcal{Q})\omega
    =-(2\mathcal{L}_\mathcal{Q}i^n_\mathcal{Q}
    +i^3_\mathcal{Q}di^{n-2}_\mathcal{Q})\omega =\cdots \\[3mm]
    = -n\mathcal{L}_\mathcal{Q}i^n_\mathcal{Q}\omega -
    i^{n+1}_\mathcal{Q}d\omega =-n\delta U (Q)-i^{n+1}_\mathcal{Q}d\omega\,.
    \end{array}
\end{equation}
Therefore,
\begin{equation}\label{h1}
\delta U(Q)=\frac 1{n+1}i^{n+1}_\mathcal{Q}d\omega =0\,.
\end{equation}
On the other hand, for an exact $\omega=d\theta$ we have
\begin{equation}\label{h2}
    U(Q)=i^n_\mathcal{Q}d\theta
    =i^{n-1}_\mathcal{Q}(d\,i_\mathcal{Q}
    +\mathcal{L}_\mathcal{Q})\theta =\cdots
    =n\mathcal{L}_\mathcal{Q}i^{n-1}_\mathcal{Q}\theta=\delta (n
    i^{n-1}_\mathcal{Q}\theta)\,.
\end{equation}
Together, Rels. (\ref{h1}) and (\ref{h2}) imply the natural
homomorphism $h: H^n_{d}(M)\to UH^n_\delta (M)$ from $n$'th group
of the De Rham cohomology to $n$'th group of $\delta$-cohomology.
This motivates to consider the factor group
\begin{equation}\label{}
\widetilde{UH}_{\delta}(M)= UH_{\delta}(M)/h(H_d(M))\,,
\end{equation}
i.e. equivalence classes of scalar $\delta$-cocycles modulo those
coming from  De Rham's cohomology classes. In the next subsection
we construct a large number of universal cocycles irreducible to
the De Rham ones.

\subsection*{Principal series of the characteristic classes.}

Let $M$ be a smooth manifold endowed with a homological vector
field $\mathcal{Q}$ (not necessarily Hamiltonian) and a symmetric
connection $\nabla$. Using these data, we define the following
pair of $(1,1)$-tensor fields:
\begin{equation}\label{bb}
\Lambda =(\nabla_i \mathcal{Q}^j)\,, \qquad
R=(R_k^l)=\left(-\frac12(-1)^{\epsilon_j}\mathcal{Q}^j\mathcal{Q}^iR_{ijk}^l\right)\,,
\end{equation}
$R_{ijk}^i(x)$ being the curvature tensor of $\nabla$. The tensors
$\Lambda$ and $R$, treated at each point of $M$ as odd and even
matrices respectively, will play the role of the building blocks
for the \textit{principal series} of universal cocycles. With
account of the nilpotency condition $\mathcal{Q}^2=0$ and the
Bianchi identity for the curvature tensor, we find
\begin{equation}\label{brel}
\nabla_\mathcal{Q} \Lambda =R-\Lambda^2 \,,\qquad
\nabla_\mathcal{Q} R=0\,.
\end{equation}

The following lemma illustrates the relevance of the  tensors
introduced to the construction of universal cocycles.

\begin{lem}  Let $\nabla$ be a flat connection, then functions
\begin{equation}\label{pser}
   \mathcal{C}_n = \mathrm{str}\,(\Lambda^{2n-1})\,,\qquad
   n=1,2,...\,,
\end{equation}
are $\delta$-closed.\end{lem}

\vspace{2mm}\noindent {\it Proof.} Since $R=0$, the relation
(\ref{brel}) yields
\begin{equation}\label{coc}
    \delta \mathcal{C}_n \equiv \mathcal{L}_\mathcal{Q}\mathcal{C}_n=\nabla_{\mathcal{Q}}\mathcal{C}_n=(2n-1)\, \mathrm{str}
    \,(\nabla_\mathcal{Q}\Lambda\cdot\Lambda^{2n-2})=-(2n-1)\,\mathrm{str}\,
    (\Lambda^{2n})=0\,,
\end{equation}
where we have used the cyclicity property of the supertrace.
$\square$

\vspace{2mm}\noindent {\it Example}. Let $\mathcal{G}$ be a Lie
algebra with a basis $e_i$ and the commutation relations
\begin{equation}\label{llbr}
    [e_i, e_j]=f_{ij}^ke_k\,.
\end{equation}
The odd vector space $\Pi \mathcal{G}$ can be endowed  with the
homological vector field
\begin{equation}\label{}
    \mathcal{Q}=\frac12 C^j C^if_{ij}^k\frac \partial{\partial
    C^k}\,,
\end{equation}
$C^i$ being odd coordinates on $\Pi\mathcal{G}$. The nilpotency
condition $\mathcal{Q}^2=0$ is equivalent to the Jacobi identity
for the Lie bracket (\ref{llbr}). Clearly, the linear space of
functions on $\Pi\mathcal{G}$ endowed with the
$\delta$-differential gives a model of the Chevalley-Eilenberg
complex for the Lie algebra $\mathcal{G}$. The cocycles
(\ref{pser}) are then nothing but the \textit{primitive elements}
of the Lie algebra cohomology
\begin{equation}\label{}
    \mathcal{C}_n (\mathcal{G})=\mathrm{tr} (\mathrm{ad}_{i_1}\cdots
    \mathrm{ad}_{i_{2n-1}})C^{i_1}\cdots C^{i_{2n-1}}\,,
\end{equation}
$\mathrm{ad}_i=(f_{ij}^k)$ being  the matrices of the adjoint
representation.

\vspace{3mm} For a general symmetric connection $\nabla$ the
formulas (\ref{brel}), (\ref{coc}) lead to the expression  $
\delta \mathcal{C}_n=\mathrm{str}\, ( R\Lambda^{2n})$  which can
be nonzero whenever $R\neq 0$. We can try to modify the expression
(\ref{pser}) by some curvature-dependent terms to restore the
closedness of $\mathcal{C}_n$. It turns out that under certain
topological restrictions  such a modification is possible indeed.
Before formulating these restrictions let us remind that $n$'th
Pontrjagin class of a tangent bundle $TM$ over an ordinary (even)
manifold $M$ can be defined as the cohomology class of the
$4n$-form $p_{n}=\mathrm{tr}({{\mathcal{R}}}^{2n})$, where
$\mathcal{R}=(\mathcal{R}_k^l)=dx^idx^j {R}_{ijk}^l$ is the
matrix-valued 2-from associated to the curvature tensor of some
symmetric connection on $M$. It easy to see that this definition
of the Pontrjagin classes is carried over to the category of
supermanifolds with the obvious replacement of the trace by the
supertrace. In perfect analogy to ordinary manifolds, one can see
that (i) the Pontrjagin classes of supermanifolds do not depend on
the choice of symmetric connection, and (ii) $(4m-2)$-forms
$\mathrm{str}(\mathcal{R}^{2m-1})$ are exact for all
$m=1,2,...\,.$

\begin{thm} For each $n=1,2,...$, there exists a unique (up to
a $\delta$-coboundary) invariant  polynomial $f_n(\Lambda,R)$ in
matrices $\Lambda$ and $R$, such that\label{th1}
\begin{equation}
f_n(\Lambda, 0)=\mathrm{str}(\Lambda^{2n-1})\,,\qquad \delta
f_n=c_n\mathrm{str}(R^n)\,,
\end{equation}
$c_n$ being some nonvanishing  constant.
\end{thm}

\begin{cor} \label{cor}If $n=2m-1$ for $m=1,2,...$, or $n=2m$ and
$2m$'th Pontrjagin class is trivial,  then there exists a
$(2n-1)$-form $\phi_n$ such that
\begin{equation}\label{chi}
\mathcal{C}_n=f_n+i_\mathcal{Q}^{2n-1}\phi_n
\end{equation}
is a $\delta$-cocycle\footnote{Informally speaking, Pontrjagin's
classes of the tangent bundle $TM$ represent obstructions for a
half of the functions (\ref{pser}) to be extendible  to
$\delta$-cocycles on a non-parallelizable $\mathcal{Q}$-manifold
$M$. }.\end{cor}

\vspace{2mm}\noindent \textit{Proof}. Under the assumptions made
\begin{equation}\label{om}
\mathrm{str}({\mathcal{R}}^n)=d\omega_n\,,
\end{equation}
for some  $(2n-1)$-form $\omega_n$. On the other hand,
\begin{equation}
i^{2n}_\mathcal{Q}\mathrm{str}({\mathcal{R}}^n) =(2n)!(-2)^n{\rm
str}(R^n)\,,
\end{equation}
and Corollary \ref{cor} readily follows from the relations
(\ref{ex}), (\ref{h1}) if one put
\begin{equation}\label{phi}
\phi_n=-\left(-\frac12\right)^n\frac{c_n\omega_n}{(2n-1)!}\,.\;\;\square
\end{equation}

\vspace{2mm}\noindent We will prove the above theorem using an
auxiliary algebraic construction which can also be understood in
terms of spectral sequences. Consider a bigraded associative
algebra
$$
\mathcal{A}=\bigoplus_{n=0,m=1}^{\infty} \mathcal{A}_{n,m}\,,
$$
freely generated by elements $a_n$, $n=0,1,...\,.$ The general
element of $\mathcal{A}$ is given by a linear combination of
homogeneous monomials
\begin{equation}\label{hommon}
A=a_{n_1}a_{n_2}\cdots a_{n_k}\,.
\end{equation}
By definition we set
\begin{equation}\label{}
    \deg_1 A = n_1+n_2+\cdots
    +n_k\,,\quad\quad \deg_2 A=k\,.
\end{equation}
It is also convenient to define the total degree $\deg
=\deg_1+2\deg_2$ and the corresponding decomposition
\begin{equation}\label{}
    \mathcal{A}=\bigoplus_{m=2}^\infty\mathcal{A}_m \,,\qquad
    \mathcal{A}_m=\bigoplus_{n+2k=m}\mathcal{A}_{n,k}\,.
\end{equation}
The algebra $\mathcal{A}$ can be endowed with a pair of nilpotent
differentials $\delta_1: \mathcal{A}_{n,m}\to \mathcal{A}_{n+1,m}$
and $\delta_2: \mathcal{A}_{n,m}\to \mathcal{A}_{n-1,m+1}$ defined
by
\begin{equation}\label{}\begin{array}{c}
    \delta_\lambda(A B)=\delta_\lambda A\cdot B+(-1)^{\deg_1 A} A\delta_\lambda B\,,\qquad \lambda=1,2\,,\\[3mm]
\delta_1 a_n=\left\{
\begin{array}{ll}
a_{n+1}\,, & \hbox{for odd $n$\,,}\\
 0 \,,&\hbox{for even $n$\,,}
\end{array}
\right.\qquad
    \delta_2 a_n=\left\{
\begin{array}{ll}
    \sum_{s=0}^{n-1}a_sa_{n-s-1}(-1)^s\,,&\hbox{for  $n > 0$\,,}\\
     0\,, & \hbox{for $n=0$\,,}
\end{array}
\right.
\end{array}
\end{equation}
for any homogeneous elements $A,B\in\mathcal{A}$. It is easy to
see that
\begin{equation}\label{}
    (\delta_1)^2=(\delta_2)^2=\delta_1\delta_2+\delta_2\delta_1=0\,,
\end{equation}
and hence $\delta=\delta_1+\delta_2: \mathcal{A}_n\to
\mathcal{A}_{n+1}$ is a nilpotent differential increasing the
total degree by 1.

Let us now extend the cochain complex $(\mathcal{A},\delta)$ to
the following one:
$$
\mathcal{A}'=\mathcal{A}\oplus \mathcal{B}\oplus \mathcal{C}\,,
$$
where $\mathcal{B}=\mathrm{span} (b_1,b_2,...)$,
$\mathcal{C}=\mathrm{span} (c_1,c_2,....)$, and
\begin{equation}\label{}
  \delta c_n=a_0^n\,, \qquad\delta b_n=\left\{%
\begin{array}{ll}
    a_{n-1}, & \hbox{for odd $n$\,,} \\
    0, & \hbox{for even $n$\,.} \\
\end{array}
\right.
\end{equation}
By definition we set $\deg b_n=n$ and $\deg c_n =2n-1$. Let
$H^n_\delta(\mathcal{A}')$, $n=1,2,...$, denote the corresponding
cohomology groups.

\begin{lem} $H^n_\delta(\mathcal{A}')\simeq
B_n=\mathrm{span}(b_n)$.\end{lem}

\vspace{3mm}\noindent
 \textit{Proof.} Define a partial homotopy operator for
 $\delta_1 $ by the rule
\begin{equation}\label{}
\begin{array}{c}
\displaystyle \delta^\ast_1 A=\frac 1{\deg_2 A}\sum_{l=1}^k
a_{n_1}\cdots a_{n_{l-1}}(\delta^\ast_1a_{n_l})a_{n_{l+1}}\cdots
a_{n_k}(-1)^{n_1+\cdots
  +n_{l-1}} \,,\\[5mm]
   \delta^\ast_1 a_n=\left\{%
\begin{array}{ll}
    a_{n-1}, & \hbox{for even $n>0$\,,} \\
    0, & \hbox{for odd $n$ or $n=0$\,,}
\end{array}
\right.
\end{array}
\end{equation}
where $A$ is given by (\ref{hommon}). One can easily see that
$(\delta_1^\ast)^2=0$ and
\begin{equation}\label{hd}
    \delta_1^\ast\delta_1+\delta_1\delta_1^\ast+\pi=1\,,
\end{equation}
$\pi$ being the canonical projection onto the subspace of
nontrivial $\delta_1$-cocycles
$\mathcal{A}_{0,\bullet}=\mathrm{span}(a_0, a_0^2,...)$.

Now we introduce the (twisting) operator
$G=1+\delta_1^\ast\delta_2: \mathcal{A}\to \mathcal{A}$. Since
$\delta_1^\ast\delta_2$ decreases the first degree by 2, the
operator $G$ is invertible. One can check that
\begin{equation}\label{}
    \delta|_\mathcal{A} =G^{-1}\delta_1 G +\pi\delta_2\,.
\end{equation}
Let us define $\delta^\ast=G^{-1}\delta_1^\ast G: \mathcal{A}\to
\mathcal{A}$. From (\ref{hd}) it then follows that
\begin{equation}\label{hdd}
    (\delta\delta^\ast+\delta^\ast\delta)a=a-\pi(G+\delta_2\delta^\ast)a\,,\qquad
    \forall\, a\in \mathcal{A}\,.
\end{equation}

Consider a general $\delta$-cocycle  $z=a+b+c$, where $a\in
\mathcal{A}$, $b\in \mathcal{C}$, $c\in \mathcal{C}$. Applying
$\delta^\ast$ to the equation $\delta z=0$ and using  (\ref{hdd}),
we obtain
\begin{equation}\label{}
\delta^\ast \delta(a+b+c)=a-\delta\delta^\ast a
-\pi(G+\delta_2\delta^\ast)a +\delta^\ast\delta b=0\,,
\end{equation}
and hence $a$ is cohomologous to $-\delta^\ast\delta b$.
Substituting $a=-\delta^\ast\delta b$ to the cocycle condition
$\delta z=0$ and using (\ref{hdd}),  we find
\begin{equation}\label{}
\pi(G+\delta_2\delta^\ast)\delta b+\delta c=0\,.
\end{equation}
Since the restriction $\delta|_\mathcal{C}$ is bijective, we can
write $c=-\delta^{-1}\pi(G+\delta_2\delta^\ast)\delta b$.
Gathering all together, we finally get $z=A b$, where
\begin{equation}\label{a}
A=1-\delta^\ast\delta-\delta^{-1}\pi(G+\delta_2\delta^\ast)\delta\,.
\end{equation}
It remains to note that the operator $A$ is invertible and
preserves the total degree. $\square$

\vspace{2mm}

\noindent Now the proof of Theorem \ref{th1} immediately follows
from the obvious identifications:
\begin{equation}\label{}
b_n=\frac 1n\Lambda^n\,,\quad a_n=R\Lambda^n\,,\quad c_n =
i^{2n-1}_\mathcal{Q}\phi_n\,,\quad \delta =\nabla_\mathcal{Q} \, ,
\end{equation}
and from the cyclicity property of the supertrace. Making use of
the operator (\ref{a}) gives us explicit expressions for the
cocycles $\mathcal{C}_n$, provided the corresponding Pontrjagin's
class (if any) is trivial. For a few first cocycles we find
\begin{equation}\label{}
\begin{array}{l}
    \mathcal{C}_1=\mathrm{str} (\Lambda) +\psi_1\,,\\[3mm]
    \mathcal{C}_2= \mathrm{str}(\Lambda^3+3R\Lambda) +\psi_3\,,\\[3mm]
\mathcal{C}_3=\mathrm{str}(\Lambda^5+5R\Lambda^3+10R^2\Lambda)+\psi_5
    \,,\\[3mm]
\mathcal{C}_4=\mathrm{str}(\Lambda^7+7R\Lambda^5+14R^2\Lambda^3+7R\Lambda
R\Lambda^2
    +35 R^3\Lambda)+\psi_7\,,
    \end{array}
\end{equation}
where $\psi_{n}=i^{2n-1}_\mathcal{Q}\phi_n$. In what follows we
will refer to $\chi_n=[\mathcal{C}_n]$, defined by (\ref{chi}), as
the principal series of characteristic classes.

\rem The form $\phi_n$, entering the definition of
$\mathcal{C}_n$, is determined by Rels. (\ref{om}), (\ref{phi}) up
to the adding any closed $(2n-1)$-form. This ambiguity, however,
does not effect on the characteristic class
$\chi_n=[\mathcal{C}_n]$ being considered as an element of
$\widetilde{UH}_\delta(M)$.

\rem As is seen, the class $\mathcal{C}_1$ is nothing but the
divergence of the homological vector field and thereby it can be
equivalently defined in terms of a nowhere vanishing density
$\rho$ instead of the connection:
\begin{equation}\label{mc}
    \mathcal{C}_1=\mathrm{div}_\rho \mathcal{Q}=\rho^{-1}\partial_i(\rho
    \mathcal{Q}^i)\,.
\end{equation}
If $\rho'$ is another density on $M$, then
$\mathcal{C}{}'_1=\mathrm{div}_{\rho'}\mathcal{Q}$ is cohomologous
to $\mathcal{C}_1$:
\begin{equation}\label{}
    \mathcal{C}{}'_1-\mathcal{C}_1 = \delta f\,,\qquad
    f=\mathrm{ln}(\rho'/\rho)\,.
\end{equation}
In particular, this class appears to be trivial for any gauge
system $(M,\{\cdot\,,\,\cdot\},Q)$ associated with an even
symplectic structure: the divergence of the Hamiltonian vector
field $\mathcal{Q}$ vanishes identically when it is evaluated
w.r.t. the Liouville volume form.

Meanwhile, for an odd Poisson bracket this class can be
nontrivial. For instance, with every Poisson manifold $(N,\pi)$,
$\pi\in \wedge^2TN$ being a Poisson bivector, one can associate a
gauge system $(\Pi T^\ast N, \{\,\cdot,\cdot\}, Q)$ w.r.t. the
canonical Poisson bracket on $\Pi T^\ast N$ and
\begin{equation}
Q=\frac12\pi ^{ij}(x)x^\ast_ix^\ast_j\,.
\end{equation}

As is known \cite{KhV},  every density $\sigma$ on $N$ corresponds
to the density $\rho=\sigma^2$ on $\Pi T^\ast N$. Using this
special density one can see that $\chi_1$ is proportional to the
modular cocycle of a Poisson manifold \cite{W}:
\begin{equation}\label{mcl}
    \chi_1=2[\sigma^{-1}\partial_i(\sigma\pi^{ij})x_j^\ast]\,.
\end{equation}
Vanishing of this class is known to be the necessary and
sufficient condition for the Poisson manifold $(N,\pi)$ to admit a
continuous trace density.

Having in mind this example, $\mathcal{C}_1$ is referred to as a
\textit{modular class} of a gauge system
$(M,\{\,\cdot,\cdot\},Q)$.

\begin{thm} \label{ind}The characteristic classes $\chi_n$ do not
depend on the choice of the symmetric connection
$\nabla$.\end{thm}

\vspace{2mm} \noindent \textit{Proof.} Given two symmetric
connections $\nabla$ and $\nabla '$, define the one-parametric
family of connections $\nabla^{(t)}=(1-t)\nabla+t\nabla'$, $t\in
\mathbb{R}$, such that $\nabla^{(0)}=\nabla$ and
$\nabla^{(1)}=\nabla '$. Consider the manifold $\tilde{M}=M\times
\mathrm{R}^{1,1}$, where $\mathrm{R}^{1,1}$  is a linear
superspace with one even coordinate $t\in \mathbb{R}$ and one odd
coordinate $\theta$. The homological vector field $\mathcal{Q}$
and the connection $\nabla^{(t)}$ are extended from $M$ to
$\tilde{M}$ as follows:
\begin{equation}\label{}
    \tilde{Q}=Q+\theta\partial_t\,,\qquad
    \tilde{\nabla}_X=\nabla^{(t)}_X\, \quad
   \tilde{ \nabla}_{\partial_t}=\partial_t\,,\quad
   \tilde{\nabla}_{\partial_\theta}=\partial_\theta\,,
    \end{equation}
where $\partial_t=\partial/\partial t$,
$\partial_\theta=\partial/\partial\theta$, and $X\in
\mathrm{Vect}(M)$. The matrices (\ref{bb}) take the form
\begin{equation}\label{mat}
    \tilde{\Lambda}=\left(%
\begin{array}{ccc}
  \Lambda^{(t)} & | & 0 \\
  --& |& -- \\
  0 & | & \begin{array}{ll}
    0 & 1 \\
    0 & 0 \\
  \end{array} \\
\end{array}%
\right)\,,\qquad \tilde{R}=\left(%
\begin{array}{ccc}
  R^{(t)} + \theta \Psi & | & 0 \\
  ---- & | & -- \\
  0 & | & \begin{array}{ll}
    0 & 0 \\
    0 & 0 \\
  \end{array} \\
\end{array}%
\right)\,,
\end{equation}
where $\Lambda^{(t)}$ and $R^{(t)}$ are constructed from
$\nabla^{(t)}$ and $\mathcal{Q}$ by the same formula (\ref{bb}),
and
\begin{equation}\label{}
    \Psi=(\mathcal{Q}^k\Delta\Gamma_{kj}^i)\,,\qquad \Delta\Gamma
    =\nabla'-\nabla\,.
\end{equation}
The principal cocycles associated to $\tilde{M},
\tilde{\mathcal{Q}}$ and $\tilde{\nabla}$ are given by
\begin{equation}\label{}
    \tilde{\mathcal{C}}_n=f_n (\Lambda^{(t)}, R^{(t)}+\theta\Psi)+\psi^{(t)}_{2n-1}=\mathcal{C}^{(t)}
    +\theta \mathrm{str}(\Psi F_n)\,,
\end{equation}
where
\begin{equation}\label{}
    \mathcal{C}^{(t)}=f_n(\Lambda^{(t)},R^{(t)})+\psi^{(t)}_{2n-1}\,,\qquad
    F_n(\Lambda^{(t)},R^{(t)})=(F^j_{ni})=({\partial f_n}/{\partial
    R^{(t)i}_j})\,.
\end{equation}
By construction,
\begin{equation}\label{}
    \mathcal{L}_\mathcal{\tilde{Q}}\tilde{\mathcal{C}}_n=\mathcal{L}_\mathcal{Q}\mathcal{C}_n^{(t)}+
    \theta [\partial_t\mathcal{C}^{(t)}_n -\mathcal{L}_\mathcal{Q} \mathrm{str}(\Psi
    F_n)]=0\,,
\end{equation}
and hence
\begin{equation}\label{}
    \mathcal{L}_\mathcal{Q}\mathcal{C}_n^{(t)}=0\,,\quad\partial_t\mathcal{C}_n^{(t)}=
    \mathcal{L}_\mathcal{Q}\mathrm{str}(\Psi F_n)\,.
\end{equation}
Integrating the last equation with respect to $t$ from $0$ to $1$,
we finally get
\begin{equation}\label{}
\mathcal{C}^{(1)}_n-\mathcal{C}^{(0)}_n=\delta \int_0^1
\mathrm{str}(\Psi F_n)dt\,,
\end{equation}
i.e., the cocycles associated to different connections belong to
the same class of $\delta$-cohomology. $\square$

\section{Cohomological operations}\label{COP}
As we have mentioned in Sec.\ref{cgs}, every tensor operation in
$\mathcal{T}(M)$ induces an operation in the $\delta$-cohomology
group $H_\delta(\mathcal{T}(M))$. In particular, scalar cocycles
of any gauge system $(M,\{\,\cdot,\,\cdot\}, Q)$ form a Poisson
algebra with respect to the point-wise multiplication and taking
the Poisson bracket. It is shown below that the latter operation
turns out to be trivial when applied to characteristic classes;
instead, the space of all universal cocycles carry a natural Lie
algebra structure with respect to another Lie bracket:
\begin{equation}\label{lbr}
    [U,V]=U\circ V - (-1)^{(\epsilon(U)+\epsilon(Q))(\epsilon(V)+\epsilon(Q))}V\circ
    U\,,
\end{equation}
where
\begin{equation}\label{circ}
   \left. U\circ V=\frac{d}{d\mu} V(Q+\mu U(Q))\right|_{\mu=0}\,,
   \qquad \epsilon(\mu)=\epsilon(Q)+\epsilon(U)\,.
\end{equation}
Notice the parity shift in the definition of the commutator of two
universal cocycles. In particular,
\begin{equation}\label{}
    \epsilon ([U,V])=\epsilon(U)+\epsilon (V)+\epsilon(Q)\,.
\end{equation}
The Jacobi identity for the commutator (\ref{lbr}) readily follows
from the definition above. As is seen the commutator of two
universal cocycles is a local map $[U,V]: C^{\infty}(M)\to
C^{\infty}(M)$. So, it remains to check that $\delta([U,V])=0$.
The last fact can be established as follows.

To each universal cocycle $U(Q)$ we assign a flow
$Q(\mu)=\varphi_\mu^U(Q)$: the evolution with respect to the
``time'' $\mu$ is described by the nonlinear PDE
\begin{equation}\label{pde}
\frac {\partial Q}{\partial \mu }=U(Q)-\frac12\mu
[U,U](Q)\,,\qquad \epsilon(\mu)=\epsilon(Q)+\epsilon(U)\,.
\end{equation}
Here we are interested in solutions given by formal power series
in $\mu$ with coefficients in $C^{\infty}(M)$. Clearly, given
$U(Q)$, there is a unique $Q(\mu)\in C^{\infty}(M)[[\mu]]$
satisfying (\ref{pde}) and the boundary condition $Q(0)=Q$. In
particular, for the odd $\mu$ the general solution reads
\begin{equation}\label{}
    Q(\mu)=Q +\mu U(Q)\,,
\end{equation}
and we see that $\{Q(\mu),Q(\mu)\}=0$. For the even $\mu$ the
commutator $[U,U]$, entering the r.h.s. of (\ref{pde}), vanishes
identically and we can write
\begin{equation}\label{}
\{\varphi_\mu^U(Q),\varphi_\mu^U(Q)\}=\varphi_{\mu}^U(\{Q,Q\})=\{Q,
Q\}+\int_0^\mu \varphi_\nu^U(\{U,Q\})d\nu =0\,.
\end{equation}
In either case, the flow $\varphi_\mu^U$ respects the master
equation. In geometric terms, one can think of $U(Q)$ as a vector
field on the space $C^\infty(M)$ (perhaps nonintegrable when
$[U,U]\neq 0$), which is tangent to the ``surface'' of all
solutions to the master equation $\{Q,Q\}=0$.

Consider now the composition of two flows. Using the definition of
$\circ$-product (\ref{circ}), we can write
\begin{equation}\label{}
    Q(\mu,\nu)=\varphi^V_\nu(\varphi^U_\mu(Q))=Q + \mu U+\nu V+
    \mu\nu V\circ U + O(\mu^2,\nu^2)\,.
\end{equation}
Since
\begin{equation}\label{}
    0=\{Q(\mu,\nu),Q(\mu,\nu)\}=
    2\mu\nu [(-1)^{\epsilon(\mu)}\{V,U\}+
    (-1)^{\epsilon(\mu)+\epsilon(\nu)}\{Q, V\circ U\}]
    +O(\mu^2,\nu^2)\,,
\end{equation}
we have
\begin{equation}\label{}
    \{V,U\}= \delta ((-1)^{\epsilon(\nu)+1} V\circ U )\,.
\end{equation}
Taking into account the symmetry property of the Poisson bracket
(\ref{PB}), we finally get
\begin{equation}\label{}
    \delta ([V,U])=0\,.
\end{equation}
Thus, the commutator of two universal cocycles is an universal
cocycle again. As a byproduct of this observation we have proved
the following statement

\begin{thm} The characteristic classes of any gauge system Poisson
commute.\end{thm}

\vspace{2mm}

\noindent It is instructive to compute the commutator of two
universal cocycles when one of them is a coboundary. We have
\begin{equation}\label{rell}
    \begin{array}{c}
       [U,\delta V]=U\circ\{Q,V\}-(-1)^{(\epsilon (U)+\epsilon(Q))
       (\epsilon(V)+\epsilon(Q)+1)}\{Q,V\}\circ U=\\[3mm]
       = \{U,V\}-(-1)^{(\epsilon (U)+\epsilon(Q))
       (\epsilon(V)+\epsilon(Q)+1)}\{Q,V\}\circ U +(-1)^{\epsilon(U)+\epsilon(Q)}\{Q,U\circ
       V\}=\\[3mm]
       =-(-1)^{(\epsilon(V)+\epsilon(Q)+1)(\epsilon(U)+\epsilon(Q)+1)}(\mathcal{L}_\mathcal{V}U)(Q)
       +(-1)^{\epsilon(U)+\epsilon(Q)}\delta(U\circ V)\,,\\
    \end{array}
\end{equation}
where $\mathcal{V}=\{V,\cdot\}$ is the Hamiltonian vector field
corresponding to  $V\in C^\infty(M)$, and
$(\mathcal{L}_\mathcal{V}U)(Q)$  is the Lie derivative of the
polydifferential operator $U$ acting on $Q$. Although the last
expression  need not be a $\delta$-coboundary in general, it is
not easy to come with a counterexample. Indeed, every cocycle of
the principal series is determined by a polydifferential operator
$\mathcal{C}_n(\pi,\nabla)(Q)$ constructed from the Poisson
bivector $\pi$ and the connection $\nabla$ by means of tensor
operations. Then
\begin{equation}\label{}
(\mathcal{L}_\mathcal{V}\mathcal{C}_n)(Q)=\mathcal{C}_n(\mathcal{L}_\mathcal{V}\pi,\nabla)(Q)
+\mathcal{C}_n(\pi,\mathcal{L}_\mathcal{V}\nabla)(Q)\,.
\end{equation}
The first term vanishes because $\mathcal{V}$ is a Hamiltonian
vector field, while the second term is a $\delta$-coboundary in
consequence of Theorem \ref{ind}.

If now $c=i_\mathcal{Q}^n\omega$ is the $\delta$-cocycle,
associated to a closed $n$-form $\omega \in \Omega^n(M)$, then one
can check that
\begin{equation}\label{}
    (\mathcal{L}_\mathcal{V}c)(Q)=
    i_\mathcal{Q}^{n}d i_\mathcal{V}\omega=\delta
    (ni_\mathcal{Q}^{n-1}i_\mathcal{V}\omega)\,,
\end{equation}
and we come with a coboundary again.

>From Rel. (\ref{rell}) also follows that the commutator of two
coboundaries is always a coboundary. For if $U=\delta W$ we have
\begin{equation}\label{}
    (\mathcal{L}_\mathcal{V}\delta W)(Q) =
    \delta[(-1)^{\epsilon(V)+\epsilon(Q)+1}(\mathcal{L}_\mathcal{V}W)(Q)]\,.
\end{equation}
Let $B\in \mathbb{L}$ denote the subalgebra of
$\delta$-coboundaries in the Lie algebra $\mathbb{L}$ of universal
cocycles, let $F\subset C^{\infty}(M)$ denote the supercommutative
subalgebra of functions generated by $\{\mathcal{C}_n\}$ and let
$L\in \mathbb{L}$ denote the Lie subalgebra generated by all the
elements of $F$. In view of the Jacobi identity for the Lie
bracket (\ref{lbr}) and the Leibnitz rule for the Lie derivative
\footnote{Notice that the adjoint action of the Lie algebra
(\ref{lbr}) does not differentiate the point-wise product of
universal cocycles.} we arrive at the following

\begin{thm} The subalgebra $L\subset \mathbb{L}$ belongs to the
normalizer of $B$ in $\mathbb{L}$ and therefore the Lie algebra
structure on $L$ is descended to its  $\delta$-cohomology space.
\end{thm}

\vspace{3mm} Commuting cocycles of the principal series one can
generate new universal cocycles involving higher derivatives of
the homological potential $Q$.

\section{Applications and interpretations of the characteristic classes}\label{apl}

As the global geometric properties of a gauge system are embodied
in the corresponding characteristic classes, the latter can find
applications every time when the global geometry becomes important
for the study of the system. Here we address two types of such
problems: the anomalies, and the theory of foliations.

It is a common knowledge  that the anomalies appearing in the
quantum field theory have a topological nature (see
e.g.\cite{ZJ}). In a wide sense, the term \textit{anomaly} means
breaking of a classical gauge symmetry upon quantization. In
practice, the anomalies manifest themselves as nontrivial
BRST-cocycles in the ghost number 1 or 2 (depending on which
formalism, Lagrangian BV or Hamiltonian BFV-BRST, is used) and
present cohomological obstructions to solvability of the quantum
counterparts to the classical master equations. In this context,
the very existence of the nontrivial cocycles in the ghost number
1 or 2 provides an indirect evidence for possible anomalies in the
respective gauge theory. In this Section we show that the modular
class of the Lagrangian gauge system appears to be the first
obstruction to the existence of a \textit{quantum master action}.
Examining the same problem for the quantum BRST charge we arrive
at a new universal cocycle having the ghost number 2 and depending
on a K\"ahler metric of the phase space to be quantized. This
analysis exhausts all the possible anomalies at the level of the
first quantum correction.

Here we also discuss the applications of our characteristic
classes to theory of (singular) foliations and give an explicit
example of a regular foliation with a nontrivial modular class.

\subsection{Anomalies in the BV-formalism.}
The quantum master action $S$, governing the quantum dynamics of a
gauge model, is defined on the same odd Poisson manifold
$(M,\{\,\cdot,\cdot\,\})$ as the classical one (see Example
\ref{lagr}), but obeys the \textit{quantum master equation}
\begin{equation}\label{qmeq}
\{S,S\}=2i\hbar\Delta S\,,
\end{equation}
$\hbar$ being the Plank constant. Here $\Delta: C^{\infty}(M)\to
C^{\infty}(M)$ is a second order differential operator, called the
\textit{odd Laplacian}, defined by
\begin{equation}\label{Lap}
\Delta f=\frac12{\rm div}_\rho\, X_f\,,
\end{equation}
where $X_f=\{f,\,\cdot\,\}$ is the Hamiltonian vector field
associated to $f\in C^\infty(M)$ and $\rho$ is a nowhere vanishing
density on $M$. It is possible to choose $\rho$ in such a way that
one can make $\rho=1$ in a neighbourhood of each point on $M$ by
an appropriate choice of Darboux coordinates for the odd
symplectic structure (this unimodularity condition is always
assumed satisfied in the BV-theory). Such a density is called
\textit{normal}, and for each normal density $\rho$ one has
\begin{equation}\label{norm}
    \Delta^2=0\,.
\end{equation}
For example, any  density $\rho=\sigma^2$, obtained by squaring a
density  $\sigma$ on the Lagrange submanifold $\Pi\mathcal{E}$, is
automatically normal.

In the Darboux coordinates the odd Laplacian was first introduced
in the works \cite{BV1},\cite{BV2} for purposes of the Lagrangian
quantization. The covariant definition (\ref{Lap}), involving a
density function, was given in  \cite{S1}, \cite{S2}, \cite{Kh}.
Various properties of odd Laplacians have been further studied and
systematized in \cite{KhV}.

The fundamental property of the odd Laplacian is that it
differentiates the odd Poisson bracket:
\begin{equation}\label{dpr}
    \Delta \{f,g\}=\{\Delta f,g\}+(-1)^{\epsilon(f)+1}\{f,\Delta
    g\}\,.
\end{equation}
This relation can be easily derived from another important formula
expressing the odd bracket via the odd Laplacian:
\begin{equation}\label{}
    \Delta (f\cdot g)=\Delta f\cdot g
    +(-1)^{\epsilon(f)}\{f,g\}+(-1)^{\epsilon(f)}f\cdot \Delta
    g\,.
\end{equation}

One can regard the quantum master equation (\ref{qmeq}) as a
one-parametric deformation of the classical one  (\ref{BVme}) with
the Plank constant being the deformation parameter. This suggests
to look for the solution in the form of a formal power series in
$\hbar$:
\begin{equation}\label{exp}
C^{\infty}(M)[[\hbar]]\ni S=S_0 + \hbar S_1+\hbar^2S_2+\cdots
\end{equation}
In view of (\ref{norm}) and (\ref{dpr}) the quantum master
equation is the Maurer-Cartan type equation and thereby it is
algebraically consistent. The local solvability of Eq.(\ref{qmeq})
was proven in Ref. \cite{BV3}. The existence of the globally
defined solution is a more difficult question due to possible
cohomological obstructions. Indeed, substituting the expansion
(\ref{exp}) into (\ref{qmeq}) one gets the following chain of
equations:
\begin{equation}\label{chain}
\begin{array}{l}
  \{S_0,S_0\}=0 \,,\\[3mm]
\displaystyle   \{S_0,S_1\}=i\Delta S_0\,,\\[3mm]
   \displaystyle\{S_0,S_n\}=i\Delta S_{n-1}+\sum_{k=1}^{n-1}\{S_k,S_{n-k}\}\,,\quad n\geq 2\,.
\end{array}
\end{equation}
The first equation identifies $S_0$ as a classical master action.
Then the rest equations take the cohomological form $\delta S_n =
B_n(S_0,...,S_{n-1})$. By induction in $n$, one can see that $B_n$
is $\delta$-closed, provided $S_0, ..., S_{n-1}$ obey the first
$(n-1)$'th equations of (\ref{chain}). Thus the existence problem
for a solution to the quantum master equation appears to be
equivalent to vanishing of the sequence of the $\delta$-cohomology
classes $[B_n]$; in so doing, $n$'th cohomology class is defined,
provided all the previous classes vanish\footnote{In algebraic
terms, the obstruction to extendibility of an $n$'th order
solution to $(n+1)$'th order one is represented by $n$'th Massey
bracket $[S_0,....,S_0]$ constructed using the Poisson bracket and
the Laplacian.}. In particular, the second equation in
(\ref{chain}) expresses the triviality of the modular class
$\chi_1=[\Delta S_0]$ associated to  the classical gauge system
$(M,\{\,\cdot,\cdot\,\},S_0)$. For the general  gauge theory
specified by the classical master action (\ref{ma}) we have the
explicit formula
\begin{equation}\label{s0}
    \chi_1 =
\left[C^\alpha(\sigma^{-1}\nabla_A(\sigma
    R^A_\alpha)+T_{\alpha\beta}^\beta) +
    \frac12\sigma^{-1}C^\beta C^\alpha \nabla_B (\sigma E_{\alpha\beta}^{AB})\phi^\ast_A+\cdots\right ]\,,
\end{equation}
where $\nabla$ is a connection on $\mathcal{E}\to N$ and $\sigma$
is a density on $N$.

It is appropriate mention that in the field-theoretical context
the modular class as well as the other characteristic classes are
not well-defined objects. The matter is that actual computations
of these classes for local functionals, like the master action,
will result in singular expressions proportional to the ``value''
of the Dirac delta-function at zero, $\delta(0)$. Nonetheless,
dividing formally by these infinite overall constants one can
obtain well-defined $\delta$-cocycles. A more rigourous way to
handle these infinities is to apply a suitable regularization
scheme (see, for instance, \cite{JST}).

\subsection{Anomalies in the Hamiltonian BRST-BFV formalism.}
In this case, the quantum master equation reads
\begin{equation}\label{meqqq}
    \frac12[\Omega,\Omega]=\Omega\ast \Omega=0\,,
\end{equation}
where $\ast$ stands for a associative product in the algebra
$C^{\infty}(M)[[\hbar]]$ of symbols of operators associated to a
Poisson manifold $(M,\{\,\cdot,\cdot\,\})$. As for any element of
this algebra, the quantum BRST charge is supposed to be given by a
formal power series in $\hbar$:
\begin{equation}\label{omexp}
    C^{\infty}(M)[[\hbar]]\ni \Omega
    =\Omega_0+\hbar\Omega_1+\hbar^2\Omega_2+\cdots\,.
\end{equation}
Once the second group of De Rham's cohomology $H^2(M)$ is
nontrivial there are infinitely many inequivalent quantizations
($\ast$-products). We refer to \cite{BCG}, \cite{Fedosovbook} for
the classification and the explicit description of inequivalent
$\ast$-products on symplectic manifolds.

It is generally accepted that the consistent quantization of
field-theoretical models should rely on the Wick symbols of
quantum observables (the representation of creation-annihilation
operators, in the simplest case). This implies the phase space of
the model $M$ to be endowed with an integrable complex structure
$J=\delta x^iJ_i^j\partial_j$ compatible with the given Poisson
structure in the sense that
\begin{equation}\label{}
    J_k^i\pi^{kl}J_l^j=\pi^{ij}\,,\qquad \nabla_k\pi^{ij}=0\,.
\end{equation}
Here $\pi^{ij}$ is the nondegenerate Poisson bivector and $\nabla$
is the symmetric connection constructed by the K\"ahler metric
$g^{ij}=J^i_k\pi^{kj}$.

The standard $\ast$-product of the Wick type \cite{DLSh},
\cite{KSh} has the following structure:
\begin{equation}\label{}
A\ast B=\sum_{n=0}^\infty \hbar^n C_n(A,B) \,,\qquad A, B\in
C^{\infty}(M)[[\hbar]]\,,
\end{equation}
where $C_n(A,B)$ is a sequence  of bilinear differential operators
the first three of which are
\begin{equation}
C_0(A,B)=A\cdot B \,,\qquad C_1(A,B)=
\frac{(-1)^{\epsilon_1}}2\lambda^{ij}\nabla_i A\nabla_j B\,,\qquad
C_2(A,B)=\frac
{(-1)^{\epsilon_2}}4\lambda^{ij}\lambda^{kl}\nabla_i\nabla_k
A\nabla_j\nabla_l B\,.
\end{equation}

$$
\epsilon_1=\epsilon_j(\epsilon(A)+\epsilon_i)\,,\qquad
\epsilon_2=(\epsilon_j+\epsilon_l)(\epsilon(A)+\epsilon_i)
+\epsilon_j\epsilon_l+\epsilon_l\epsilon_k+\epsilon_k\epsilon_i\,.
$$
The Hermitian form $\lambda^{ij}=g^{ij}+i\pi^{ij}$ on the
complexified cotangent bundle of $M$ is called the Wick tensor.
Substituting the expansion (\ref{omexp}) into (\ref{meqqq}) we get
the chain of equations
\begin{equation}\label{brst3}
    \{\Omega_0,
    \Omega_n\}=-\sum_{k+l+m=n+1}C_k(\Omega_l,\Omega_m)\,,\qquad
    l,m < n\,.
\end{equation}
The equation $\{\Omega_0,\Omega_0\}=0$, appearing at the first
order in $\hbar$, is recognized as the master equation for the
classical BRST charge $\Omega_0$ (cf. (\ref{bfvme})). As in the
previous example, the system of equations (\ref{brst3}) has a
cohomological form: in order for $n$'th-order solution to exist,
the r.h.s. should be a $\delta$-coboundary, while \textit{a
priory} one can only ensure its $\delta$-closedness (The latter
fact follows from the Jacobi identity for the $\ast$-commutator).
In particular, for the first quantum correction $\Omega_1$ to the
classical BRST charge we get
\begin{equation}\label{jl}
    \delta \,\Omega_1=C_2(\Omega_0,\Omega_0)=\frac12{\rm str}\,
    (J\Lambda^2)\,,
\end{equation}
where as before
$\delta=\{\Omega_0,\cdot\,\}=\mathcal{Q}^i\partial_i$ and
$\Lambda=(\nabla_i\mathcal{Q}^j)$. The cocycle
$C_2(\Omega_0,\Omega_0)$ has the ghost number 2 and, as we will
see, may happen to be nontrivial. Indeed, using the relations
(\ref{brel}) and the compatibility condition $\nabla J=0$, one can
easily find that
\begin{equation}\label{dcob}
    \delta \,{\rm str}\,(J\Lambda )={\rm str}(JR)-{\rm str}
    (J\Lambda^2)\,.
\end{equation}
This shows that $C_2(\Omega_0,\Omega_0)$ is cohomologous to
$(1/2)i^2_\mathcal{Q}\rho\,$, where
\begin{equation}\label{}
\rho=\frac 12dx^i
dx^j\mathcal{R}_{ijk}^lJ_l^k(-1)^{\epsilon_k}\,,\qquad d\rho=0\,,
\end{equation}
is the Ricci form of the K\"ahler manifold $M$. The De Rham class
of a Ricci 2-form is known to be proportional to the first Chern
class of a complex manifold. When the latter does not vanish the
$\delta$-cohomology class of $i^2_\mathcal{Q}\rho$ can be
non-vanishing as well giving the first obstruction for the quantum
BRST charge $\Omega$ to exist. However, this conclusion  refers
only to the chosen $\ast$-product. To overcome the obstruction one
can fine tune the $\ast$-product by deforming the original Poisson
structure in the ``direction'' of the Ricci form,
\begin{equation}\label{shift}
\pi\to \pi'=(\pi^{-1}+\hbar\rho)^{-1}=\pi-\hbar\,
\pi\rho\pi+O(\hbar^2)\,,
\end{equation}
and define then the new $\ast$-product w.r.t. the modified Wick
tensor $\lambda'= J\pi'+i\pi'$. In such a way we pass from the
standard $\ast$-product of the Wick type, invariantly specified by
the characteristic class $\mathrm{cl}(\ast)=\pi^{-1}$, to an
inequivalent star-product $\ast'$ with
$\mathrm{cl}(\ast')=\pi^{-1}+\hbar[\rho]$. For more details on the
equivalence problem of the Wick type star-products see Refs.
\cite{BCG}, \cite{DLSh}, \cite{KSh}. It is straightforward to
check that the shift (\ref{shift}) completes the r.h.s. of
(\ref{jl}) to the $\delta$-coboundary (\ref{dcob}). As a result we
can write the first-order correction to the classical BRST charge
as
\begin{equation}\label{evlap}
    \Omega_1 = \frac12{\rm str}\,(J\Lambda )+\delta \psi=\frac12\Delta \Omega_0 +\delta\psi\,,\qquad \forall
    \psi\in C^{\infty}(M)\,,
\end{equation}
$\Delta=g^{ij}\nabla_i\nabla_j$ being the Laplace operator
associated to the K\"ahler metric.

We are lead to conclude  that, choosing an appropriate Wick
$\ast$-product, one can always find a formal solution to the
quantum master equation at least in the first order in $\hbar$.
This holds true whenever one deals with finite dimensional
manifolds.  In the field theory, however, the Laplace operator is
known to be an ill-defined object: acting on a local functional it
gives singular expressions in prefect analogy to the previous case
of the odd Laplacian. Of course, one can try to remove these
singularities by means of an appropriate regularization but this
can (and very so often does) destroy the exactness of the cocycle
(\ref{dcob}). Thus, in the infinite dimensional setting this
cocycle may happen to be nontrivial even though one can write it
as the formal coboundary (\ref{dcob}) \footnote{A good analogy is
the integrability problem for a closed 1-form $\theta$ on a
multi-connected manifold $M$: there always exist a potential $f$,
such that $df=\omega$, if one admits multi-valued (= ill-defined)
functions on $M$.}. The conformal anomaly in the string theory
\cite{BE}, \cite{BFa} is a typical example of this kind.

\subsection{Characteristic classes of  foliations}\label{hclf}
Given a regular foliation $\mathcal{F}$ on an ordinary manifold
$N$, denote by $\mathcal{E}\subset TN$ the subbundle given by the
union of tangent spaces to leaves of the foliation $\mathcal{F}$
(so-called, tangent bundle of a foliation). Since $\mathcal{E}$ is
integrable, the inclusion map $A: \mathcal{E}\to TM$ defines the
Lie algebroid $\mathcal{E}$ over $N$. The regularity of $F$
implies the anchor $A$ to be injective. Thus, there is the
one-to-one correspondence between the categories of regular
foliations and injective Lie algebroids. To any Lie algebroid, in
turn, we can associate the gauge system (see Example \ref{liealg})
and define the corresponding characteristic classes. When the Lie
algebroid comes from a regular foliation, these  characteristic
classes  can be attributed to the foliation itself.

In particular, the modular class of the foliation  $\mathcal{F}$
is determined by the cocycle
\begin{equation}\label{mclass}
    \mathcal{C}_1 = (\sigma^{-1}\nabla_i(\sigma
    A_\alpha^i)+T_{\alpha\beta}^\beta)C^\alpha\,,
\end{equation}
where $\sigma$ is a density on $N$ and $\nabla$ is a connection on
$\mathcal{E}\to N$ (cf. (\ref{s0})).

The nontriviality of the modular class can be illustrated by the
following example. Choose a discrete subgroup $\Gamma\subset
SL(2,\mathbb{R})$ in such a way that the right quotient
$N=SL(2,\mathbb{R})/\Gamma$ to be a compact manifold. (It is well
known that such subgroups do exist.) Let $\{e_i\}$ be Weyl's basis
in the space of right invariant vector fields on
$SL(2,\mathbb{R})$:
\begin{equation}\label{sl2}
[e_{-1},e_1]=2e_0\,,\quad [e_0, e_{1}]=e_{1}\,,\quad
[e_0,e_{-1}]=-e_{-1}\,.
\end{equation}
The canonical projection $\varphi: SL(2,\mathbb{R})\to
SL(2,\mathbb{R})/\Gamma = N$ allows one to descend  the vector
fields $e_i$ to $N$, so that the resulting vector fields $e'_i$
obey the same commutation relations (\ref{sl2}). Explicitly,
\begin{equation}\label{comrel}
\mathrm{d}\varphi_g e_i(g)=e'_i(\varphi(g))\,,\qquad \forall g\in
SL(2,\mathbb{R})\,.
\end{equation}
(The vector fields $e_i$ and $e'_i$ are said to be
$\varphi$-related.) The pair $(e'_0,e'_1)$ generates the action of
the Borel subgroup $B\in SL (2,\mathbb{R})$ on $N$ and hence
defines a two dimensional foliation $\mathcal{F}$. The
corresponding Lie algebroid structure is assigned to the trivial
vector bundle $B\times N\to N$.

Notice that the dual to the 3-vector $e'_{-1}\wedge e'_0\wedge
e'_{1}$ is the $SL(2,\mathbb{R})$-invariant volume form $\sigma$
on $N$. Therefore,  $\mathrm{div}_\sigma e_i'=0$.  Choosing a flat
connection on $B\times N\to N$ and substituting the structure
constants (\ref{sl2}) into (\ref{mclass}) we get a nontrivial
cocycle $\mathcal{C}_1=C^0$. For if $\mathcal{C}_1$ is trivial one
could find a function $f\in C^{\infty}(N)$ such that $e'_0f=1$.
But every function on a compact manifold has at least two critical
points at which $e'_0f=0$. Thus the modular class of the foliation
at hands is nontrivial.

\section{Conclusion}
Let discuss some of the paper results and related issues.

In this paper we use a broad concept of the gauge systems going
beyond the traditional physical applications and covering a
diversity of other geometric structures like Lie and Courant
algebroids, etc. Making use of this universal framework, we study
the problem of constructing topological invariants which can be
uniformly attributed to every gauge system, characterizing the
global geometric properties of the system. As any gauge system is
described in this framework by a nilpotent operator, called a
homological vector field, the most simple and natural invariants
to look for are those belonging to the respective cohomology and
constructed from the derivatives of the homological vector field
itself. We refer to these invariants as the characteristic classes
having in mind the analogous constructions for the vector bundles
and foliations and, as we have shown, this is more then just an
analogy (see Sec. \ref{hclf}). A sequence of characteristic
classes, called the principal series, has been explicitly
constructed involving the first derivatives of the homological
vector field.

The universal concept of the gauge system offers several
advantages over the conventional approaches which are specific for
every particular type of the gauge models. Even in the well
studied case of Lie algebroids this concept seems useful,
providing concise and geometrically transparent depiction for the
respective structures, their morphisms and cohomological
invariants. Besides, it provides a simple description for certain
cohomological operations we have found in this paper. Namely, the
space of characteristic classes is endowed with a Lie bracket
operation which is not induced by the original Poisson bracket
(the latter is proved to be trivial when applied to characteristic
classes). By making use of this Lie bracket one can derive new
characteristic classes depending on higher derivatives of the
homological vector field. A general method still remains to be
worked out for constructing higher derivative invariants which are
not deducible from the principal characteristic classes by means
of these cohomological operations. The first derivative invariants
appear to be sufficient, however, for many important physical
applications like the computation of one-loop anomalies in the
quantum field theory.

As is seen, many known types of the gauge systems admit this
uniform homological description, though the examples we listed in
the paper by no means exhaust all the diversity of the geometric
structures which can be treated in this framework. In this
homological language, the distinctions between different types of
gauge systems are encoded in the ghost number and parity grading
in the target space. By scrutinizing all the admissible
combinations of ghost number and parity, one can perhaps reveal
some new meaningful geometrical constructions emerging from the
master equation.

Finally, let us note that the classical gauge systems can be in
themselves described in terms of the homological vector field
alone, with no reference to the homological potential and the
Poisson bracket; and the same holds true for the  principal series
of characteristic classes. Where the Poisson structure becomes
relevant it is in the problem of deforming the classical gauge
systems. In the case of the even Poisson structure, the powerful
tool for that is provided by the deformation quantization
technique which can be applied for this purpose just by replacing
the Poisson square of the homological potential with a respective
$\ast$-square. In the case of the odd Poisson structure, one can
use the BV quantum master equation which is also recognized as an
efficient tool for deforming algebraic structures \cite{St2}.
Although the both methods first appeared as quantization schemes
for gauge systems, they can help to construct the deformations
which are not necessarily quantum by their nature.

\appendix
\section{Tensor calculus on supermanifolds}
We list here the notation and conventions we use in the paper,
which mostly coincide with those adopted in \cite{Leit}. Given a
smooth supermanifold $M$, denote by $C^\infty(M)$ the
supercommutative algebra of smooth functions on $M$. We fix the
basic field to be $\mathbb{R}$, though all the formulae are still
valid for an arbitrary algebraic field of characteristic 0.

\subsection{Tensor fields on supermanifolds.} Recall that
an endomorphism $D: C^{\infty}(M)\to C^{\infty}(M)$ is said to be
a differentiation of the Grassman parity $\epsilon(D)$ if
\begin{equation}\label{}
    D(fg)=D(f)g + (-1)^{\epsilon(D)\epsilon(f)}f D(g)\,.
\end{equation}
Similar to the case of ordinary manifolds, the vector fields on
$M$ can be identified with elements of the Lie superalgebra
$\mathrm{Der}(M)$ of all the differentiations of $C^\infty(M)$.
With this geometrical interpretation the $C^{\infty}(M)$-module
$\mathrm{Der}(M)$ will be denoted $\mathrm{Vect}(M)$. In each
coordinate chart $(U, x^i)$, the restriction
$\mathrm{Vect}(U)=\mathrm{Vect}|_U(M)$ is a free
$C^\infty(M)$-module with the basis $\partial_i$, so that
\begin{equation}\label{}
X(f)=X^i\partial_if \,,
\end{equation}
$f,\, X^i \in C^{\infty}(U)$. Hereafter
$\partial_i=\partial/\partial x^i$ stands for the left derivative
on $C^{\infty}(U)$. On each nonempty intersection $U\cap U'$ of
two coordinate charts the basis vector fields are related to each
other by the rule
\begin{equation}\label{}
    \frac{\partial}{\partial x^i}=\left(\frac{\partial x'^{j}}{\partial x^i}\right)\,\frac{\partial}{\partial x'^j}\,.
\end{equation}
The superspace
$\mathrm{Vect}(M)=\mathrm{Vect}_0(M)\oplus\mathrm{Vect}_1(M)$ is
decomposed in the direct sum of two subspaces constituted,
respectively, by even and odd vector fields:
\begin{equation}\label{}
\begin{array}{rcl}
  \mathrm{Vect}_0(M) \ni X & \Leftrightarrow &  \epsilon(X)=\epsilon(X^i)+\epsilon_i =
  0\,,\\[3mm]
  \mathrm{Vect}_1(M) \ni Y& \Leftrightarrow &
  \epsilon(Y)=\epsilon(Y^i)+\epsilon_i=1\,,\\[3mm]
  \forall \, i&=&1,\dots,\dim \,M\,.
\end{array}
\end{equation}
Hereafter we put $\epsilon_i=\epsilon(x^i)$. The supercommutator
of two vector fields is given by
\begin{equation}\label{}
    [X,Y]=(X^j\partial_j Y^i-(-1)^{\epsilon(X)\epsilon(Y)}Y^j\partial_jX^i)\partial_i\,.
\end{equation}

The space of covector fields $\mathrm{Covect}(M)$ is defined as
the $C^\infty(M)$-module $ \mathrm{Hom}(\mathrm{Vect}(M),
C^{\infty}(M))$. Introduce an even $\mathbb{R}$-linear operator
$\delta: C^{\infty}(M)\to \mathrm{Covect}(M)$, called
defferential, as
\begin{equation}\label{}
    \delta f(X)=X(f)\,,\quad\quad \forall f\in
    C^\infty(M)\,,\;\;\forall X\in\mathrm{Vect}(M)\,.
\end{equation}
By definition, the operator $\delta$ is a differentiation from
$C^{\infty}(M)$ to $\mathrm{Covect}(M)$, i.e. $\delta
(fg)=\delta(f)g+f\delta(g)$. In each coordinate chart $(U,x^i)$
the covector fields $\delta x^i$ form a basis in the
$C^{\infty}(U)$-module $\mathrm{Covect}(U)$, dual from the right
to the basis $\partial_i$.  The contraction (or pairing) of a
covector field $A=\delta x^iA_i(x)$ with a vector field
$X=X^i(x)\partial_i$ is the superfunction
\begin{equation}\label{}
    \langle X,A\rangle =A(X)=X^iA_i\,.
\end{equation}

Tensoring $C^\infty (M)$-modules $\mathrm{Vect}(M)$ and
$\mathrm{Covect}(M)$, we obtain the tensor algebra
$$
\mathcal{T}(M)=\bigoplus_{\epsilon,n,m}\,\mathcal{T}_\epsilon^{(n,m)}(M)\,,\quad\quad
\epsilon=0,1\,,\;\;\;\; m,n=0,1,2,...\,,
$$
where $\mathcal{T}^{(0,0)}(M)=C^{\infty}(M)$, and the general
element of $\mathcal{T}^{(n,m)}_\epsilon(M)$ reads
\begin{equation}\label{}
\begin{array}{c}
    S =\delta x^{j_1}\otimes \cdots\otimes \delta x^{j_m} S^{i_1\cdots i_n}_{j_1\cdots
    j_m}(x)\partial_{i_1}\otimes\cdots\otimes
    \partial_{i_n}\,,\\[3mm]
    \epsilon =\epsilon(S)=\epsilon(S^{i_1\cdots i_n}_{j_1\cdots j_m})+
    \sum_{k=1}^n\epsilon_{i_k} +\sum_{l=1}^m \epsilon_{j_l}\,.
\end{array}
\end{equation}
The tensor product $S\otimes T$ is defined using the following
conventions:
\begin{equation}\label{conv}
    \delta x^i f =(-1)^{\epsilon(f)\epsilon_i}f\delta x^i\,,\quad
    \delta x^i\otimes
    \partial_j=(-1)^{\epsilon_i\epsilon_j}\partial_j\otimes\delta
    x^j\,,\quad f\partial_i=(-1)^{\epsilon(f)\epsilon_i}\partial_i\cdot f \,,\quad\forall f\in
    C^{\infty}(M)\,.
\end{equation}
According to the definition, $\epsilon(S\otimes
T)=\epsilon(S)+\epsilon(T)$.

Identifying the space $\mathcal{T}^{(1,1)}(M)$ with the space of
endomorphisms of $C^{\infty}(M)$-module $(\mathrm{Covect}(M))$, we
turn the former to a $\mathbb{Z}_2$-graded unital algebra over
$C^{\infty}(M)$. The product of two elements $S, T\in
\mathcal{T}^{(1,1)}(M)$ is the $(1,1)$-tensor $ST$ whose
components are
\begin{equation}\label{}
(ST)^i_j= S_j^kT_k^i\,.
\end{equation}
Clearly, $\epsilon(ST)=\epsilon(S)+\epsilon(T)$. The contraction
of vector and covector fields gives rise to  a linear homomorphism
$\mathrm{str}: \mathcal{T}^{(1,1)}(M)\to C^\infty(M)$:
\begin{equation}\label{}
    \mathrm{str}\,(S) = \sum (-1)^{(\epsilon(S)+1)\epsilon_i}S^i_i \,.
\end{equation}
This homomorphism,  called \textit{supertrace}, is characterized
by the property of vanishing on supercommutators, i.e.
\begin{equation}\label{}
    \mathrm{str}\, ([S,T])= \mathrm{str}\,(ST)-(-1)^{\epsilon(S)\epsilon(T)}\mathrm{str}\,(TS)=0\,, \quad\quad
    \forall\,
    S,T \in \mathcal{T}^{(1,1)}(M)\,.
\end{equation}

\subsection{The Lie derivative.} The space $\mathcal{T}(M)$ can
be endowed with a module structure over the Lie superalgebra
$\mathrm{Vect}(M)$. The corresponding homomorphism $\mathcal{L} :
\mathrm{Vect}(M)\to \mathrm{End}(\mathcal{T}(M))$ is known as the
\textit{Lie derivative}. For any $f\in C^{\infty}(M)$ and
$X,Y\in\mathrm{Vect}(M)$ we set
\begin{equation}\label{}
    \mathcal{L}_X f=X(f)=(\delta f)(X)\,,\quad\quad
    \mathcal{L}_XY=[X,Y]\,,
\end{equation}
where $\mathcal{L}_X=\mathcal{L}(X)$. Requiring the operator
$\mathcal{L}_X$ to be compatible with the pairing, i.e.
\begin{equation}\label{}
\mathcal{L}_X\langle Y,A\rangle=\langle \mathcal{L}_X
Y,A\rangle+(-1)^{\epsilon(X)\epsilon(Y)}\langle Y,\mathcal{L}_X
A\rangle \,,
\end{equation}
and to satisfy the Leibnitz rule
\begin{equation}\label{}
\mathcal{L}_X (S\otimes T)=\mathcal{L}_X S\otimes T +
(-1)^{\epsilon(X)\epsilon(S)}S\otimes \mathcal{L}_X T\,,
\end{equation}
one extends the action of $\mathcal{L}_X$ to the whole tensor
algebra $\mathcal{T}(M)$. For example, the Lie derivative of a
$(1,1)$-tensor $S=(S^i_j)$ is the $(1,1)$-tensor $\mathcal{L}_XS$
with the components
\begin{equation}\label{}
(\mathcal{L}_XS)^i_j=(-1)^{\epsilon(X)\epsilon_j}X^k\partial_kS^i_j+\partial_jX^kS_k^i
-(-1)^{\epsilon(X)\epsilon(S)} S_j^k\partial_kX^i\,.
\end{equation}
It follows from the definition that
\begin{equation}\label{}
    [\mathcal{L}_X,\mathcal{L}_Y]=\mathcal{L}_X\mathcal{L}_Y-
    (-1)^{\epsilon(X)\epsilon(Y)}\mathcal{L}_Y\mathcal{L}_X=\mathcal{L}_{[X,Y]}\,.
\end{equation}

\subsection{ The affine connection.} This can be defined as an even
map $\nabla: \mathrm{Vect}(M)\otimes \mathrm{Vect}(M)\to
\mathrm{Vect}(M)$, which is $C^\infty(M)$-linear in the first
argument, additive in the second argument, and such that
\begin{equation}\label{}
\nabla_X(fY)=X(f)Y+(-1)^{\epsilon(f)\epsilon(X)}f\nabla_X Y\,,
\end{equation}
for any $f\in C^{\infty}(M)$, $X, Y \in \mathrm{Vect}(M)$. Here
$\nabla_X(Y)=\nabla (X,Y)$. The operator $\nabla_X $ is called the
\textit{covariant derivative} along the vector field $X$. Setting
$\nabla_X f=X(f)$, postulating the compatibility of $\nabla_X$
with the pairing,
\begin{equation}\label{}
    \nabla_X \langle Y,A\rangle=\langle\nabla_X Y, A\rangle+(-1)^{\epsilon(X)\epsilon(Y)}\langle Y,\nabla_X
    A)
\end{equation}
as well as the Leibnitz rule
\begin{equation}\label{}
    \nabla_X(S\otimes T)=\nabla_X S\otimes T +
    (-1)^{\epsilon(S)\epsilon(X)}S\otimes \nabla_X T\,,
\end{equation}
one extends the action of $\nabla_X$ from $\mathrm{Vect}(M)$ to
the whole tensor algebra $\mathcal{T}(M)$. In particular, for a
$(1,1)$-tensor $S$ we have
\begin{equation}\label{}
    (\nabla_X S)^i_j= (-1)^{\epsilon(X)\epsilon_j} X^k\partial_k S^i_j-(-1)^{\epsilon(X)\epsilon_j}
    X^k\Gamma_{kj}^n S_n^i
+(-1)^{\epsilon(X)(\epsilon(S)+\epsilon_i)}
    S_j^n X^k\Gamma_{kn}^i\,,
\end{equation}
where $\Gamma_{ij}^k(x)= \delta x^k(\nabla
(\partial_i,\partial_j))$ are Christoffel's symbols. An affine
connection $\nabla$ is said to be \textit{symmetric} if
\begin{equation}\label{}
    \nabla_X Y-(-1)^{\epsilon(X)\epsilon(Y)}\nabla_Y X=[X,Y]\,.
\end{equation}
In terms of local coordinates this means that
\begin{equation}\label{}
    \Gamma_{ij}^k (x)=(-1)^{\epsilon_i\epsilon_j}\Gamma_{j\,i}^k(x)\,.
\end{equation}

The curvature of an affine connection is given by the
supercommutator of the corresponding covariant derivatives. For a
symmetric connection $\nabla$ we have
\begin{equation}\label{}
    [\nabla_i,\nabla_j](\partial_k)=\nabla_i\nabla_j \partial_k -
    (-1)^{\epsilon_i\epsilon_j}\nabla_j\nabla_i \partial_k =R_{ijk}^l(x)\partial_l\,,
\end{equation}
where $\nabla_i=\nabla_{\partial_i}$ and the components of the
curvature tensor read
\begin{equation}\label{}
    R_{ijk}^l=\partial_i\Gamma_{jk}^l-(-1)^{\epsilon_i\epsilon_j}\partial_j\Gamma_{ik}^l
    +(-1)^{(\epsilon_j+\epsilon_k+\epsilon_m)\epsilon_i}\Gamma_{jk}^m\Gamma_{im}^l
    -(-1)^{(\epsilon_k+\epsilon_m)\epsilon_j}\Gamma_{ik}^m\Gamma_{jm}^l\,.
\end{equation}

\subsection{The parity reversion operation.}
 Let $V=V_0\oplus V_1$ be a $\mathbb{Z}_2$-graded linear space
(superspace) such that $\epsilon (v)=0$ and $\epsilon(u)=1$ for
any $v\in V_0$ and $u\in V_1$. Reversing parities of all the
homogeneous elements of $V$ we get the new superspace $\Pi(V)$
with $\epsilon(v)=1$ and $\epsilon(u)=0$. Clearly,
$\Pi(\Pi(V))=V$. The homomorphism $\Pi: V\to \Pi(V)$ is called the
\textit{parity reversion}. Let $T(V)=\bigoplus V^{\otimes n}$
denote the tensor algebra of the superspace $V$. The
\textit{symmetric algebra} $S(V)$ of $V$ is defined as the
quotient $T(V)/I$, where $I\subset T(V)$ is the two-sided ideal
generated by elements of the form $x\otimes y-(-1)^{\epsilon
(x)\epsilon(y)}y\otimes x$. Thus $S(V)$ is a supercommutative
associative algebra freely generated by elements of $V$. The
\textit{exterior algebra} ${E}(V)$ of $V$ is then defined as the
symmetric algebra of $\Pi(V)$.

\subsection{Differential forms.} Using the parity reversion operation,
the exterior algebra of differential forms $\Omega (M)$ can be
defined as the supercommutative algebra ${E}(\mathrm{Covect}(M))$.
By definition, $\Omega(M)=\bigoplus \Omega^n(M)$, where
$\Omega^0(M)=C^{\infty}(M)$,
$\Omega^1(M)=\Pi(\mathrm{Covect}(M))$, and the subspace
$\Omega^n(M)$ of $n$-forms, $n\geq 1$, is multiplicatively
generated by elements of $\Omega^1(M)$.  For any $f\in
C^{\infty}(M)$ denote by $df$ the differential 1-form $\Pi(\delta
f)\in \Omega^1(M)$. Clearly, $\epsilon(df)=\epsilon (f)+1$.  In
each coordinate chart $(U,x^i)$ the forms $dx^i$ constitute a
basis in the $C^{\infty}(U)$-module $\Omega^1(U)$, so that any
$n$-form $\omega\in \Omega^n(U)$ can be written as
\begin{equation}\label{}
 \omega   =dx^{i_1}\cdots dx^{i_n}\omega_{i_1\cdots
 i_n}(x)\,,\qquad \omega_{i_1\cdots i_n}\in C^{\infty}(U)\,.
\end{equation}
According to the definition above (cf. (\ref{conv}))
\begin{equation}\label{}
    dx^idx^j=(-1)^{(\epsilon_i+1)(\epsilon_j+1)}dx^jdx^i\,,\quad
    dx^i f(x)=(-1)^{\epsilon(f)(\epsilon_i+1)}f(x)dx^{i}\,,\quad
    \forall \,f\in C^{\infty}(U)\,.
\end{equation}

The \textit{exterior derivative} is the unique differentiation $d:
\Omega^n(M)\to \Omega^{n+1} (M)$ possessing the properties:
    \begin{enumerate}
        \item $\epsilon(d)=1$;
        \item  if $f\in C^{\infty}(M)$, then $df=\Pi(\delta f)$;
        \item $d^2=0$.
    \end{enumerate}
In terms of local coordinates it is given by  $d=dx^i\partial_i$.

The \textit{interior multiplication} of a form  by a vector field
$X\in \mathrm{Vect(M)}$ is the unique differentiation $i_{X}:
\Omega^n(M)\to \Omega^{n-1}(M)$ obeying conditions:
\begin{enumerate}
    \item $\epsilon(i_X)=\epsilon(X)+1$;
    \item $i_{X}f=0\,,\quad \mathrm{for}\;\forall\,f\in
    C^{\infty}(M)$;
    \item $i_X(df)=(-1)^{\epsilon(X)}Xf$.
\end{enumerate}
The following identities hold true:
\begin{equation}\label{}
[i_X,i_Y]=0\,,\quad [d,i_X]=\mathcal{L}_X\,,\quad
[\mathcal{L}_X,i_Y]=(-1)^{\epsilon(X)\epsilon(Y)}i_{[X,Y]}\,,
\end{equation}
where $\mathcal{L}_X$ is the Lie derivative along the vector field
$X\in \mathrm{Vect}(M)$.

\end{document}